\documentclass[preprint,11pt]{elsarticle}

\usepackage{graphicx}
\usepackage[utf8]{inputenc}
\usepackage{multirow}
\usepackage{comment}
\usepackage{mdwlist}
\usepackage{amsfonts}
\usepackage{ulem}
\usepackage{amsmath} 
\usepackage{amsthm}
\usepackage{amssymb}
\usepackage{fancyhdr}
\newtheorem{lemma}{Lemma}
\newtheorem{theorem}{Theorem}
\usepackage{titlesec}
\usepackage{indentfirst}
\usepackage{booktabs}
\usepackage{verbatim}
\usepackage{color}
\usepackage{latexsym}
\usepackage{comment}
\usepackage{amscd}
\usepackage{esvect}
\usepackage{graphicx}
\usepackage{upgreek}
\usepackage{caption}   
\usepackage[a4paper]{geometry}
\usepackage{mathrsfs}
\usepackage[numbers]{natbib}
\usepackage{float}
\usepackage{lipsum}
\usepackage[export]{adjustbox} 
\usepackage{subfiles}
\usepackage{subfigure}
\usepackage{bm}
\usepackage{epstopdf}
\newcommand{\e}{\varepsilon}
\newcommand{\bx}{\boldsymbol{x}}

\newcommand{\bv}{\boldsymbol{v}}
\newcommand{\blue}{\textcolor{black}}
\newcommand{\red}{\textcolor{black}}

\begin{document}

\title{Asymptotic-preserving neural networks for the semiconductor Boltzmann equation and its application on inverse problems}
\author{Liu Liu, Yating Wang, Xueyu Zhu, Zhenyi Zhu}

\begin{abstract}
In this paper, we develop the Asymptotic-Preserving Neural Networks (APNNs) approach to study the forward and inverse problem for the semiconductor Boltzmann equation. The goal of the neural network is to resolve the computational challenges of conventional numerical methods and multiple scales of the model. To guarantee the network can operate uniformly in different regimes,
it is desirable to carry the Asymptotic-Preservation (AP) property in the learning process. In a micro-macro decomposition framework, we design such an AP formulation of loss function. The convergence analysis of both the loss function and its neural network is shown, based on the Universal Approximation Theorem and hypocoercivity theory of the model equation. We show a series of numerical tests for forward and inverse problems of both the semiconductor Boltzmann and the Boltzmann-Poisson system to validate the effectiveness of our proposed method, which addresses the significance of the AP property when dealing with inverse problems of multiscale Boltzmann equations  especially when only sparse or partially observed data are available. 
\end{abstract}

\maketitle

\section{Introduction}

Kinetic equations have been widely used in many areas such as rarefied gas, plasma physics, astrophysics, semiconductor device modeling, social and biological sciences \cite{Semi-Book}. They describe the non-equilibrium dynamics of a system composed of a large number of particles and bridge atomistic and continuum models in the hierarchy of multiscale modeling. The Boltzmann-type equation, as one of the most representative models in kinetic theory, provides a power tool to describe molecular gas dynamics, radiative transfer, plasma physics and polymer flow \cite{A15}. They have significant impacts in designing, optimization, control and inverse problems. For example, it can be used in the design of semiconductor devices, topology optimization of gas flow channel, or risk management in quantitative finance \cite{CS21}. Many of these applications often  require finding unknown or optimal parameters in the Boltzmann-type equations or mean-field models \cite{AHP15,Caflisch, CFP13, Cheng11, Lai}. 

The linearized Boltzmann equation, in particular the semi-conductor Boltzmann equation that we will study in this work, has significant applications in the semiconductor device modeling \cite{Ansgar}. 
In this work, we will study the forward problem of approximating the solution of the model and the inverse problem which aims to infer the collision cross-section in the semiconductor Boltzmann equation, with available full or partial observation data given as measurable density quantity. 

Modeling and predicting the evolution of multiscale systems such as the Boltzmann-type equations have always been challenging, which often requires sophisticated knowledge of numerical methods and labor-intensive implementation, in addition to the prohibitive costs due to the well-known curse of dimensionality. In real-life applications that usually involve the scarcity of available data, multiple scales and uncertainties, solving physical problems with missing or incomplete initial or boundary conditions via traditional numerical approaches thus becomes highly impractical. This motivates researchers to develop data-driven models and methods \cite{EW21} in the recent decade. 

Machine Learning, or Deep neural networks (DNNs) in particular, have gained increasing interest in approximating the solutions partial differential equations (PDE) due to its universal approximation property and its ability to handle high-dimensional problems. Many studies have been done to employ DNN for solving deterministic or parameterized PDEs, and have shown remarkable promise in various applications. Among those, Physics informed neural network (PINN) \cite{PINN2, raissi2019physics} is one of the most famous approach which incorporates the available physical laws and limited data, such as boundary or initial conditions and the source term, to approximate the solution of the underlying PDE. The idea has been successfully applied in the simulation of many forward and inverse problems \cite{chen2020physics, lou2021physics, mao2020physics}. 

For the kinetic problems under study, another challenge is the multiple scale lying in the models, which makes it computational difficult to resolve small scales especially when the physical scaling parameter is small. Asymptotic-preserving (AP) schemes, which preserve the asymptotic transition from one scale to another at the discrete level, have been a popular and robust computational method in the past two decades \cite{AP-Review} for many kinetic and hyperbolic problems. The main advantage of AP schemes is that they are very efficient in all regimes--especially the hydrodynamic or diffusive regime, since they do not need to resolve the small physical parameters numerically yet can still capture the macroscopic behavior.

However, if one directly applies the standard PINNs when dealing with these multiscale model, it may lead to incorrect inferences and predictions \cite{Jin-Ma-Wu1}. This is due to the presence of small scales need to be enforced consistently during the learning process, but a standard PINN formulation only captures the solution at the leading order of the Knudsen number, thereby losing accuracy in the asymptotic limit regimes. 
To overcome this difficulty, the authors in \cite{APNN-transport} propose the Asymptotic-preserving neural networks (APNNs) to enhance the performance of standard PINN for solving multiscale linear transport equations. The loss function of APNN contains the micro-macro decomposition of the original problem and some conservation conditions, which has shown to be efficient and accurate in capturing the limiting macroscopic behavior of the solution when the scaling parameter approaches zero. For the hyperbolic type linear kinetic equations with multiple scales, there are some existing work that studied the APNN method, for example, see \cite{Guilia-Zhu, Jin-Ma-Wu2, Jin-Ma-Wu1}. 

In this work, we will show how to construct APNNs for the semiconductor Boltzmann equation and show its effectiveness in solving both forward and inverse problems of interests.  To our best knowledge, this is the {\it first} work we employ the APNN method under the micro-macro decomposition framework for the semiconductor Boltzmann equation.

Besides, inspired by existing work \cite{Zheng-Liu-Mu, Hwang} that have studied the convergence of the loss function and the neural network solution for the linear transport equation and Fokker-Planck equation. 
We show the convergence result based on the hypercoercivity analysis, that is {\it uniform} in the Knudsen number and owns an exponential decay in time. We remark that our analysis framework on the APNNs can be extended similarly to a class of linear and nonlinear kinetic models.

The rest of the paper is organized as follows. 
In Section \ref{sec:model}, we first introduce the model equations including the semiconductor Boltzmann and the Boltzmann-Poisson system, then derive a coupled system in the micro-macro decomposition framework with a formal analysis of asymptotic limit. 
In Section \ref{sec:NN-general}, we review the PINNs and introduce APNNs method for our model by constructing the loss function that preserves the AP property in the training of neural networks. 
In Section \ref{sec:analysis}, we provide the convergence analysis of both the loss function and its neural network approximated solution, based on the universal approximation theorem and hypercoercivity theory for kinetic models. 
In Section \ref{sec:Num}, a series of numerical experiments are shown to validate the effectiveness of our proposed APNN method, where both forward and inverse problems using synthetic data generated by the kinetic model. Finally, conclusions and future work will be discussed in the last section.

\section{The models and Micro-macro decomposition method}
\label{sec:model}

\noindent\textbf{The semiconductor-Boltzmann.}
We consider the semiconductor Boltzmann equation \cite{JP00} given by 
\begin{equation}
\label{Boltz-eqn}
\e \partial_t f + v \partial_x f +  \partial_x \phi\, \partial_v f = \frac{1}{\e}\mathcal{Q}(f), 
\end{equation}
where $f(t,x,v)$ is the probability density distribution for particles located at $x \in \mathcal{D}$ with velocity $v\in \mathbb{R}$. In our model, $\e$ is the Knudsen number defined as the ratio of the mean free path and the typical length scale. In our application, $\e$ varies from $O(1)$, the kinetic regime, to $\e\ll 1$, the diffusion regime.  the  $\phi(t,x)$ is external electric potential. 

The anisotropic collision operator $\mathcal{Q}$ describes a linear approximation of the electron-phonon interaction, given by 
$$  \mathcal{Q}(f)(t,x,v) = \int_{\mathbb R} \sigma(v,w) \left( M(v)f(t,x,w) - M(w)f(t,x,v) \right) dw, $$
with $M$ the normalized Maxwellian 
$ M(v) = \frac{1}{\sqrt{\pi}} e^{-v^2}$. 
Here $\sigma(v,w)$ denotes the scattering coefficient for the electron-phonon collisions. The collision frequency is defined as
$$ \lambda(v) = \int_{\mathbb{R}} \sigma(v,w)M(w) dw. $$
We refer the readers to \cite{Rode,Ansgar,Sze81} for more physical background.

Define $n_x$ the unit outward normal vector on the spatial boundary 
$\partial\mathcal{D}$. Let $\gamma = \partial\mathcal{D}\times\Omega$, then the phase boundary can be split into an outgoing boundary $\gamma_{+}$, incoming boundary $\gamma_{-}$ and a singular boundary $\gamma_0$, which are defined by
\begin{equation}
\label{gamma}
\begin{aligned}
& \gamma_{+}:= \left\{ (x,v)\in \partial\mathcal{D}\times\Omega: 
v \cdot n_x >0 \right\}, \\[6pt]
& \gamma_{-}:= \left\{ (x,v)\in \partial\mathcal{D}\times\Omega: v\cdot n_x <0 \right\}, \\[6pt]
& \gamma_0 := \left\{ (x,v)\in \partial\mathcal{D}\times\Omega: v\cdot n_x =0\right\}. 
\end{aligned}
\end{equation}
The inflow boundary condition is given by
\begin{equation} f(t,x,v) = f_{\text{BC}}(t,x,v), \qquad \text{for   }\, 
(t,x,v) \in [0,T]\times \gamma_{-}. \end{equation}
We assume the initial condition that
$$ f(t=0,x,v)=f_{\text{IC}}(x,v). $$

\noindent\textbf{The Boltzmann-Poisson system.}
In this work, we also study the {\it nonlinear} Boltzmann-Poisson system, where the electric potential $\phi(t,x)$ is solved by the Poisson equation as below: 
\begin{equation}
\label{eqn:BP}
\left\{
\begin{array}{ll}
\displaystyle\e \partial_t f + v \partial_x f + 
\partial_x \phi\, \partial_v f  = \frac{1}{\e}\mathcal{Q}(f), \\[6pt]
\displaystyle\beta \partial_{xx}\phi = \int_{\mathbb{R}}f dv - c(x), \\[6pt]
\displaystyle\phi(t,0) = 0, \qquad \phi(t,1) = V, 
\end{array}
\right.
\end{equation}
with $\beta$ the scaled Debye length, $V$ the applied bias voltage and $c(x)$ the doping profile.

%--------------------------------------------------
\subsection{The micro-macro decomposition method}

We employ the micro-macro decomposition technique \cite{MM08} and derive the micro-macro system for the semiconductor Boltzmann equation \eqref{Boltz-eqn}. 
Assume the ansatz 
\begin{equation}\label{Ans} f = \blue{\Pi f} + \e g, 
\end{equation}
where $g: = g(t, x, v)$ and the notation \blue{$\Pi f :=  \langle f \rangle\, M(v)$} is defined by  
\begin{equation*}\label{bracket}
    \langle f \rangle := \int_{\mathbb{R}} f(t,x,v)\,dv = \rho(t, x).
\end{equation*}

Inserting the ansatz \eqref{Ans} into  \eqref{Boltz-eqn}, one gets
\begin{equation}\label{PP}
\e \partial_t (\Pi f) + \e^2 \partial_t g + v\, \partial_x (\Pi f) + \e \partial_x (v g) \red{ - 2 v \rho\, \partial_x \phi \, M(v) + \e \partial_x\phi\, \partial_v g } = \mathcal{Q}(g), 
\end{equation}
where $\partial_v (\Pi f) = \rho\, \partial_v M(v) = - 2 \rho\, v M(v)$ is used. 

Take the projection operator $\Pi$ on both sides of \eqref{PP}, then 
$$ \e \partial_t (\Pi f) + \e^2 \partial_t (\Pi g) + \partial_x \rho \left(\int v M(v) dv\right) M(v) + \e \Pi \partial_x (v g) + \e \partial_x\phi\, \Pi(\partial_v g) = 0, $$
which is equivalent to
\begin{equation}\label{Macro}
\partial_t (\Pi f) + \Pi \partial_x (v g) + \red{ \partial_x\phi \, \Pi(\partial_v g)} = 0, 
\end{equation}
where $\Pi g = 0$ and $\int_{\mathbb R} v M(v) dv = 0$ is used. 
Now subtract \eqref{PP} by \eqref{Macro}, then
$$
\e^2 \partial_t g  + v \partial_x (\Pi f)  + \e (\mathbb{I} - \Pi) \partial_x (v g) + \red{ \e \partial_x \phi\,  (\mathbb{I} - \Pi) \partial_v g - 2 v \rho\, \partial_x \phi \, M(v)} = \mathcal{Q}(g), 
$$
that is
\begin{equation}\label{Micro}
\e^2 \partial_t g +  v \partial_x \rho\, M(v)  +  \red{\e (\mathbb{I} - \Pi) \left( v\, \partial_x g + \partial_x\phi\, \partial_v g \right) - 2 v \rho\, \partial_x \phi \, M(v) } = \mathcal{Q}(g).  
\end{equation}

As $\e\to 0$, equation \eqref{Macro} stays unchanged, and the microscopic equation \eqref{Micro} gives 
$$ g = \mathcal{Q}^{-1}\left( v\, \partial_x \rho\, M(v) - 2 v \rho\, \partial_x\phi\, M(v) \right) = \left( \partial_x\rho - 2 \rho\, \partial_x\phi \right) \mathcal{Q}^{-1}(v M(v)). $$ 
Plug $g$ into \eqref{Macro} and integrate over $v$, we derive the drift-diffusion limit (\cite{PP91}) which reads
\begin{equation}\label{Diffusion} \partial_t \rho = \red{ T \left( \partial_{xx}\rho - 2\partial_x (\rho\, \partial_x\phi) \right)} , \end{equation}
where in the 1D slab geometry we have $ T = \int_{\mathbb{R}} \frac{v^2}{\lambda(v)}\, M(v) dv$.  

To summarize, the coupled equations \eqref{Macro}--\eqref{Micro} for $\rho$ and $g$ solve the following system in our micro-macro decomposition framework: 
\begin{equation}
\label{MM}
\left\{
\begin{array}{ll}
\displaystyle\partial_t \rho + \partial_x \langle v g \rangle + \red{\partial_x\phi \,\langle \partial_v g \rangle} = 0, \\[6pt]
\e^2 \partial_t g +  v \partial_x \rho\, M(v)  +  \red{\e (\mathbb{I} - \Pi) \left( v\, \partial_x g + \partial_x\phi\, \partial_v g \right) - 2 v \rho\, \partial_x \phi \, M(v) } = \mathcal{Q}(g), \\[6pt]
\Pi g = 0. 
\end{array} 
\right.
\end{equation}
where the third equation is for the mass conservation for $g$, which is crucial to the training of neural networks \cite{Jin-Ma-Wu2} and will be mentioned in section \ref{sec:APNN}. 

As $\e\to 0$, according to \eqref{MM}, the limiting system for $\rho$ and $g$ satisfies
\begin{equation}
\label{limit-diff}
\left\{
\begin{array}{ll}
\displaystyle\partial_t \rho + \partial_x \langle v g \rangle + \red{\partial_x\phi \,\langle \partial_v g \rangle} = 0, \\[6pt]
\displaystyle v  \partial_x \rho \, M(v) - 2 v \rho \partial_x\phi \, M(v)  = \mathcal{Q}(g) , \\[6pt]
\displaystyle \Pi g = 0, 
\end{array} 
\right.
\end{equation}
which is {\it equivalent} to the diffusion limit  given in \eqref{Diffusion}.

%---------------------------------------------------
\section{Neural Network Approaches}
\label{sec:NN-general}
\subsection{Physics Informed Neural Networks}
\label{sec:NN}

The standard Physics-informed Neural Networks (PINNs) \cite{PINN2, raissi2019physics} incorporate residual associated with the underlying PDE, along with initial and boundary conditions or other physical properties (when applicable) into the loss function.
It adopts automatic differentiation and embeds the physical laws in addition to the data-driven terms in the loss function at a given set of points in a computational domain. 

We first review the standard PINN approach in our problem setting. For forward problems, our goal is to approximate the solution of the semiconductor Boltzmann equation \eqref{Boltz-eqn}. Suppose the neutral network has $L$ layers; the input layer takes $(t,x,v)$ and the final layer gives $f^{NN}(t,x,v;m,w,b)$ or simply $f^{NN}_{\theta}$ %\textcolor{red}{define $\theta$}
with $\theta$ representing the neural network parameters, 
as the output. The relation between the $l$-th and $(l+1)$-th layer ($l=1,2, \cdots L-1)$ is 
given by
\begin{equation}\label{two-layer}
    n_j^{(l+1)}=\sum_{i=1}^{m_{l}} w_{j i}^{(l+1)} \sigma_{l} (n_{i}^{l})+b_{j}^{(l+1)},
\end{equation} 
where $m=\left(m_{0}, m_{1}, m_{2}, \dots, m_{L-1}\right)$, $w=\left\{w_{j i}^{(k)}\right\}_{i, j, k=1}^{m_{k-1}, m_{k}, L}$ and $b=\left\{b_{j}^{(k)}\right\}_{j=1, k=1}^{m_{k}, L}$. More specifically, 
\begin{itemize}
    \item $n_i^l$: the $i$-th neuron in the $l$-th layer 
    \item $\sigma_l$: the activation function in the $l$-th layer
    \item $w_{j i}^{(l+1)}$: the weight between the $i$-th neuron in the $l$-th layer and the $j$-th neuron in the $(l+1)$-th layer
    \item $b_j^{(l+1)}$: the bias of the $j$-th neuron in the $(l+1)$-th layer
    \item $m_l$: the number of neurons in the $l$-th layer. 
\end{itemize}

For forward problems, we desire to approximate the solution of the semiconductor Boltzmann equation \eqref{Boltz-eqn} with given initial and boundary conditions. To find the optimal values for the network parameters $\theta$ that are composed of all the weights $w_{ji}$ and bias $b_j$, the neural network is trained by minimizing the following loss function 
\begin{equation}
\label{Loss-PINN}
\begin{aligned}
\mathcal{R}_{\mathrm{PINN}}^{\e}(\theta) = Loss_{\text{GE}}(\theta) + \lambda_1 Loss_{\text{BC}}(\theta) + \lambda_2 Loss_{\text{IC}}(\theta),  
 \end{aligned}
\end{equation}
where the loss for the residual of the governing equation \eqref{Boltz-eqn} given by: 
\begin{equation}\label{L-GE}
Loss_{\text{GE}}(\theta):=\int_{\mathcal{T}} \int_{\mathcal{D}} \int_{\Omega} \Big| \e \partial_t f_\theta^{\mathrm{NN}}
 +   v \partial_x f_\theta^{\mathrm{NN}} +  \partial_x \phi\, \partial_v f_\theta^{\mathrm{NN}} - \frac{1}{\e}\mathcal{Q}(f_\theta^{\mathrm{NN}})|^2 \, d\bv d\bx dt. 
\end{equation}
The loss for the boundary condition is
\begin{equation}\label{L-BC} Loss_{\text{BC}}(\theta):= \int_{\mathcal{T}}\int_{\gamma_{-}}
\left|\mathcal{B}  \left(f_{\theta}^{NN}\right) - f_{\text{BC}}\right|^2 \, ds dt,  \end{equation}
and the loss for the initial condition: 
\begin{equation}\label{L-IC} Loss_{\text{IC}}(\theta):=\int_{\mathcal{D}} \int_{\Omega}\left|\mathcal{I}\left(f_{\theta}^{NN}\right)- f_{\text{IC}} \right|^2
 \, d\bv d\bx, 
\end{equation}
where the initial and boundary conditions are incorporated into the loss function as a regularization or penalty term, 
with penalty parameters $\lambda_1, \lambda_2$ chosen for good performance. The most popular method to minimize loss function over the parameter space are stochastic gradient descent and advanced optimizers such as Adam \cite{Adam14}. After the training process and achieving the optimal set of parameter values $\theta^{\star}$ by minimizing the PINN loss
\eqref{Loss-PINN}, i.e., 
$$ \theta^{\star} = \text{argmin} \ \mathcal{R}_{\mathrm{PINN}}^{\e}(\theta), $$
the neural network surrogate $f^{NN}_{\theta^{\star}}(t,x,v)$ can be evaluated at any given point in the temporal and phase space to obtain the solution of \eqref{Boltz-eqn}.

In the context of inverse problems, the structure of the network is almost the same as the forward problem setting, except that the unknown physical parameters of interests $\xi$ are considered as learnable parameters. In this case, we aim to optimize network parameters $\theta$ and $\xi$ jointly, namely
$$ (\theta^{\star}, \xi^{\star}) = \text{argmin} \ \mathcal{R}_{\mathrm{PINN}}^{\e}(\theta, \xi), $$
with
$$
\mathcal{R}_{\mathrm{PINN}}^{\e}(\theta, \xi) = Loss_{\text{GE}}(\theta,\xi) + \lambda_1 Loss_{\text{BC}}(\theta,\xi) + \lambda_2 Loss_{\text{IC}}(\theta,\xi) + \lambda_3 Loss_{\text{Data}}(\theta, \xi),  
$$
where $Loss_{\text{GE}}$, $Loss_{\text{BC}}$ and $Loss_{\text{IC}}$ are the same as \eqref{L-GE}, \eqref{L-BC} and \eqref{L-IC}, and the data mismatch loss $Loss_{\text{Data}}$ is involved when additional measurement data are available in the inverse problems. In our problem setting, our targeted parameter of interest is the collision kernel $\sigma$. Both available observation data will be employed in our numerical experiments for inverse problems. Details of available observation data will be explained in Section \ref{sec:Num}.

%-------------------------------------------------
\subsection{The APNN framework}
\label{sec:APNN}

Due to the Knudsen number $\e$ in our model equations, the traditional PINNs ignore the asymptotic property and cannot capture the macroscopic behavior {\it uniformly} with respect to $\e$, thus leading to wrong approximations. 
In contrast, the strategy of designing Asymptotic-preserving neural networks (APNNs), first brought up by \cite{Jin-Ma-Wu2}, is to define a new loss function that can capture the correct asymptotic behavior of the model equations, especially when the physical scaling parameter is small; in other words, the loss function has the AP property. 
The diagram \ref{fig:APNN} below illustrates the idea of APNNs \cite{Jin-Ma-Wu2}:

\begin{figure}[htbp]
\centering
\includegraphics[width=0.6\textwidth]{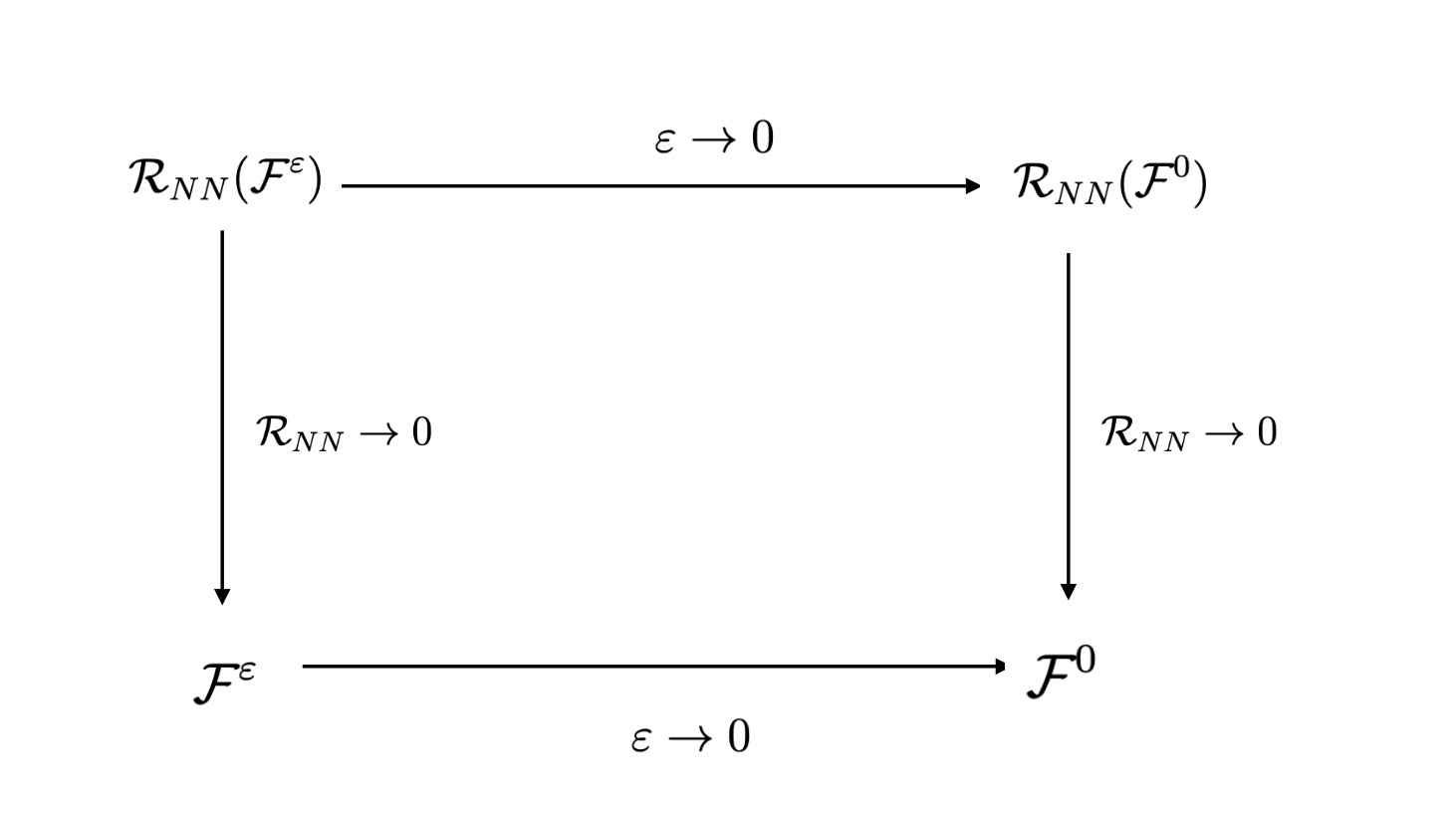}
\caption{Illustration of APNNs. }
\label{fig:APNN}
\end{figure}

Let $\mathcal{F}^{\e}$ be a multiscale model that depends on the scaling parameter $\e$. As $\e\to 0$, it converges to a reduced order limit 
$\mathcal{F}^0$. The solution of $\mathcal{F}^{\e}$ is approximated by the neural network through the imposition of residual term $\mathcal{R}_{NN}(\mathcal{F}^{\e})$, 
whose asymptotic limit is denoted by $\mathcal{R}_{NN}(\mathcal{F}^0)$ as $\e\to 0$. The neural network is called AP if $\mathcal{R}_{NN}(\mathcal{F}^0)$ is consistent
with the residual for the system $\mathcal{F}^0$. 

\textbf{Solution Representation. } We employ two deep neural networks to parameterize the two functions $\rho(t,\bx)$ and $g(t, \bx, \bv)$ in system \eqref{MM}, where the notations here are extended to the multi-dimensional spatial and velocity space. Let $\tilde{\rho}_{\theta}^{\mathrm{NN}}$ be the output of a neural network with inputs $t$ and $\bx$. Since the density $\rho$ is non-negative, we put an exponential function at the last output layer as a post-processing procedure, i.e., let
\begin{equation}
\rho_\theta^{\mathrm{NN}}(t, \bx):=\exp \left(-\tilde{\rho}_\theta^{\mathrm{NN}}(t, \bx)\right) \approx \rho(t, \bx)
\end{equation}
be the approximation for $\rho$. 

We remark that the velocity discretization for $g(t,\bx,\bv)$, where the Hermite quadrature rule is used since $\bv \in (-\infty, \infty)$. This is  standard and the same as traditional AP method studied in \cite{JP00}. Let $g(t, \bx, \bv) = \psi(t, \bx, \bv)M(\bv)$, with $M(\bv) = \frac{1}{\sqrt{\pi}} e^{-\bv^2}$ and 
\begin{equation}\label{Psi} \psi(t,\bx,\bv) = \sum_{k=0}^{N} \psi_k(t,\bx) H_k(\bv)
\end{equation}
is the Hermite expansion with $N$ being the order. 
For simplicity, we omit the $t$ and $\bx$ dependence of functions below. Denote $H_k$ as the renormalized Hermite polynomials with $H_{-1}=0$, $H_0 = 1/\pi^{1/4}$, and 
$$ H_{k+1} = v \sqrt{\frac{2}{k+1}}H_k - \sqrt{\frac{k}{k+1}}H_{k-1} \quad \text{for } k\geq 0. $$
The inverse Hermite expansion is defined by
\begin{equation}\label{I-Psi} \psi_k = \sum_{j=0}^{N_{\bv}} \psi(v_j)\, H_k(v_j)\, w_j, 
\end{equation}
where $(v_j, w_j)$ are the points and corresponding weights of the Gauss-Hermite quadrature rule. 
We then compute the collision operator $Q$ in \eqref{Boltz-eqn} as follows
$$ Q(g)(\bv) = M(\bv) \sum_{j=0}^{N_{\bv}} \sigma(\bv, v_j)\, \psi(v_j)\, w_j - \lambda(\bv) g(\bv), $$
with $\lambda(\bv) = \sum_{j=0}^{N_{\bv}} \sigma(\bv, v_j)\, w_j$. 
The derivative in $\bv$ for $\psi$ is given by
\begin{equation}
\begin{aligned}
\partial_{\bv} \psi & = \sum_{k=0}^{N} \psi_k\, \partial_{\bv} H_k(\bv) 
= \sum_{k=0}^{N} \psi_k \sqrt{2k}\, H_{k-1}(\bv) \\
& = \sum_{k=0}^{N} \sum_{j=0}^{N_{\bv}} \psi(v_j) \, H_k(v_j) w_j \sqrt{2k}\, H_{k-1}(\bv) 
 = \sum_{j=0}^{N_v} \psi(v_j)\, C_j(\bv),
\end{aligned}
\end{equation}
where $C_j(\bv) = \sum_{k=0}^{N} \sqrt{2k}\, H_k(v_j) H_{k-1}(\bv) w_j$. 

In order to obtain approximation for $g$, we let $\tilde{\psi}_\theta^{\mathrm{NN}}(t,\bx,\bv)$ be the output of a fully-connected neural network with input $t$, $\bx$ and $\bv$, then impose a post-processing step to guarantee the mass conservation in \eqref{MM}. Define 
\begin{equation}\label{g_NN}
 g_\theta^{\mathrm{NN}}(t, \bx, \bv) =
 \psi_{\theta}^{\mathrm{NN}}(t,\bx,\bv)M(\bv) := \tilde{\psi}_\theta^{\mathrm{NN}}(t,\bx,\bv)M(\bv)-\Pi\left(\tilde{\psi}_\theta^{\mathrm{NN}}(t,\bx,\bv)M(\bv)\right), 
\end{equation}
as an approximation for $g$, now 
$\Pi g_\theta^{\mathrm{NN}} = 0 $ is automatically satisfied. Moreover, $\nabla_{\bv}g_\theta^{\mathrm{NN}}$ is computed by 
\begin{equation}
    \nabla_{\bv}g_\theta^{\mathrm{NN}} = \sum_{j=0}^{N_{\bv}}\psi_\theta^{\mathrm{NN}}(t,\bx,v_j) C_j(\bv) M(\bv)-2\bv M(\bv)\psi_\theta^{\mathrm{NN}}(t,\bx,\bv) + 2\bv M(\bv) \langle \psi_\theta^{\mathrm{NN}}(t,\bx,\bv) M(\bv) \rangle.
\end{equation}
where $\nabla_{\bv}\psi_{\theta}^{NN} = \sum_{j=0}^{N_{\bv}} \psi_\theta^{\mathrm{NN}}(t,\bx,v_j)C_j(\bv)$.  

\bigskip

\textbf{APNN Loss. } For our APNN method, we propose the physics-informed loss based on  the residual for the macro-micro system \eqref{Macro}--\eqref{Micro} as the loss function: 
\begin{equation}
\label{Loss-APNN}
\begin{aligned}
\mathcal{R}_{\mathrm{APNN}}^{\e} = & \frac{1}{|\mathcal{T} \times \mathcal{D}|} \int_{\mathcal{T}} \int_{\mathcal{D}}\left|\partial_t \rho_\theta^{\mathrm{NN}}+\nabla_{\bx} \cdot\left\langle\bv g_\theta^{\mathrm{NN}}\right\rangle + \red{ \nabla_{\bx}\phi \cdot  \left\langle \nabla_{\bv} g_\theta^{\mathrm{NN}}\right\rangle } \right|^2 \, d\bx d t \\[6pt]
& + \frac{1}{|\mathcal{T} \times \mathcal{D} \times \Omega|} \int_{\mathcal{T}} \int_{\mathcal{D}} \int_{\Omega} \Big| \e^2 \partial_t g_\theta^{\mathrm{NN}}
 + \e(I-\Pi) \left(\bv \cdot \nabla_{\bx} g_\theta^{\mathrm{NN}} + \red{\nabla_{\bx}\phi \cdot \nabla_{\bv} g_\theta^{\mathrm{NN}} }\right) \\[6pt]
&   - 2\bv \cdot \nabla_{\bx}\phi\, \rho_\theta^{\mathrm{NN}}\, M(\bv)  
 + \bv \cdot \nabla_{\bx} \rho_\theta^{\mathrm{NN}} M(\bv)  - \mathcal{Q}( g_\theta^{\mathrm{NN}} ) \Big|^2 \,  d\bv d\bx d t  \\[6pt]
& + \frac{\lambda_1}{| \mathcal{T} \times \partial \mathcal{D} \times \Omega |} \int_{\mathcal{T}}\int_{\gamma_{-}}\left|\mathcal{B}  \left(\rho_\theta^{\mathrm{NN}} M(\bv) + \e g_\theta^{\mathrm{NN}}\right) - f_{\text{BC}}\right|^2 \, ds dt\\[6pt]
 &+\frac{\lambda_2}{|\mathcal{D} \times \Omega|} \int_{\mathcal{D}} \int_{\Omega}\left|\mathcal{I}\left(\rho_\theta^{\mathrm{NN}} M(\bv) +\e g_\theta^{\mathrm{NN}}\right)- f_{\text{IC}} \right|^2
 \, d\bv d\bx.
\end{aligned}
\end{equation}

As $\e\to 0$, the above first two terms arisen from the model equation lead to 
\begin{equation*}
\begin{split}
& \mathcal{R}_{\mathrm{APNN}} \text {, residual}  =  \frac{1}{|\mathcal{T} \times \mathcal{D}|} \int_{\mathcal{T}} \int_{\mathcal{D}}\left|\partial_t \rho_\theta^{\mathrm{NN}}+\nabla_x \cdot\left\langle\bv g_\theta^{\mathrm{NN}}\right\rangle + \red{ \nabla_{\bx}\phi \cdot  \left\langle \nabla_{\bv} g_\theta^{\mathrm{NN}}\right\rangle } \right|^2 \, d \bx d t \\[6pt]
& +\frac{1}{|\mathcal{T} \times \mathcal{D} \times \Omega|} \int_{\mathcal{T}} \int_{\mathcal{D}} \int_{\Omega}\left|   - 2\bv \cdot \nabla_{\bx}\phi\, \rho_\theta^{\mathrm{NN}}\, M(v) 
+ \bv \cdot \nabla_{\bx} \rho_\theta^{\mathrm{NN}}\, M(v)  -  \mathcal{Q}(g_\theta^{\mathrm{NN}})\right|^2 \, d \bv d \bx d t, 
\end{split}
\end{equation*}
which is exactly the loss for the limiting equations \eqref{limit-diff}. Thus our designed loss function for APNN satisfies the AP property that the traditional PINN fails to capture. 

\textbf{APNN Schematics. } Finally, we summarize the framework of our method for the forward problem in Figure \ref{fig:Boltzmann_framework} below. Here $\sigma(x)$ denotes the activation function. 
\begin{figure}[H]
\centerline{\includegraphics[width=1\linewidth]{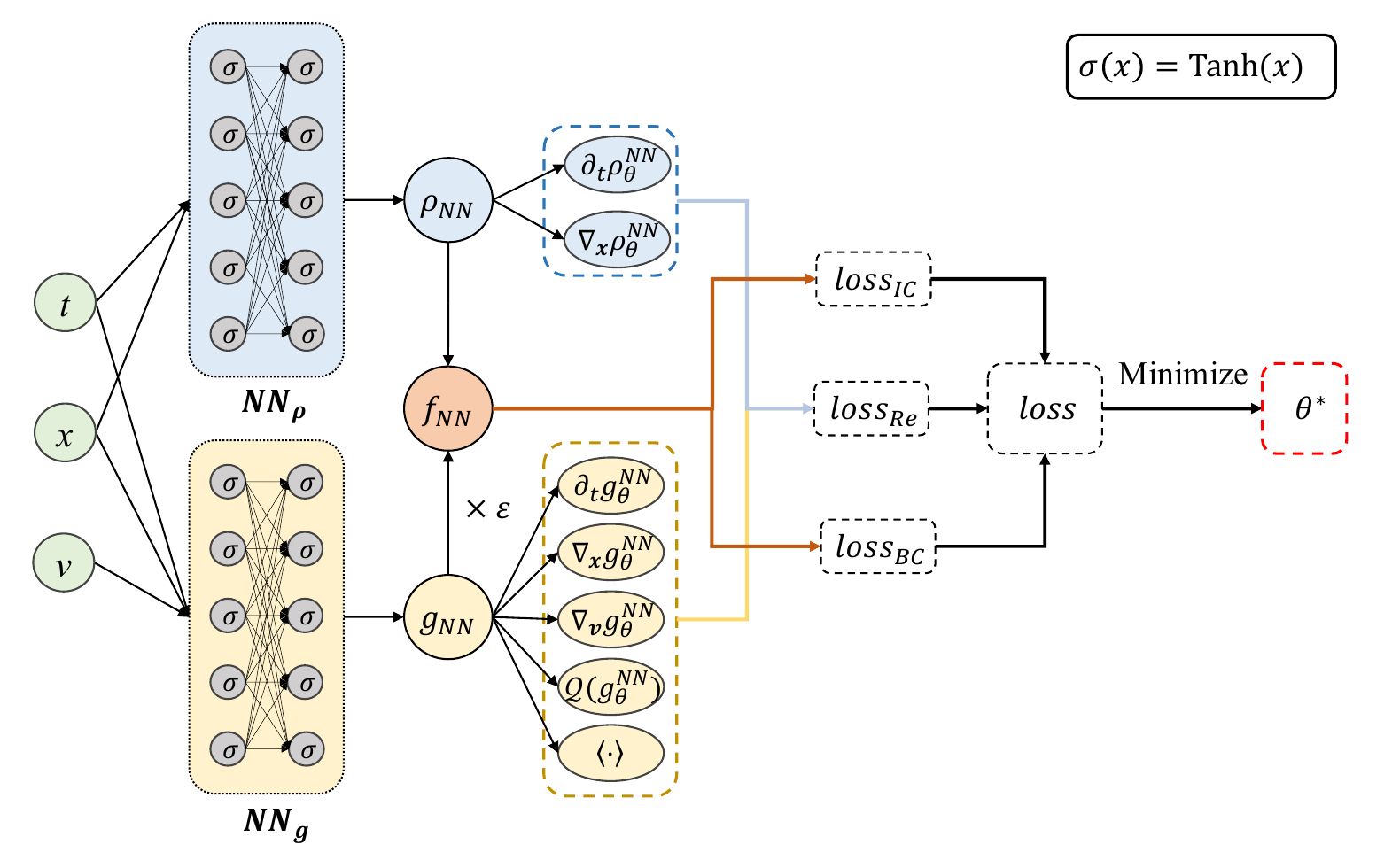}}
    \caption{APNNs framework of solving the forward problem for the semi-conductor Boltzmann-(Poisson) equation. }
    \label{fig:Boltzmann_framework}
\end{figure}

\begin{figure}[H]
\centerline{\includegraphics[width=1\linewidth]{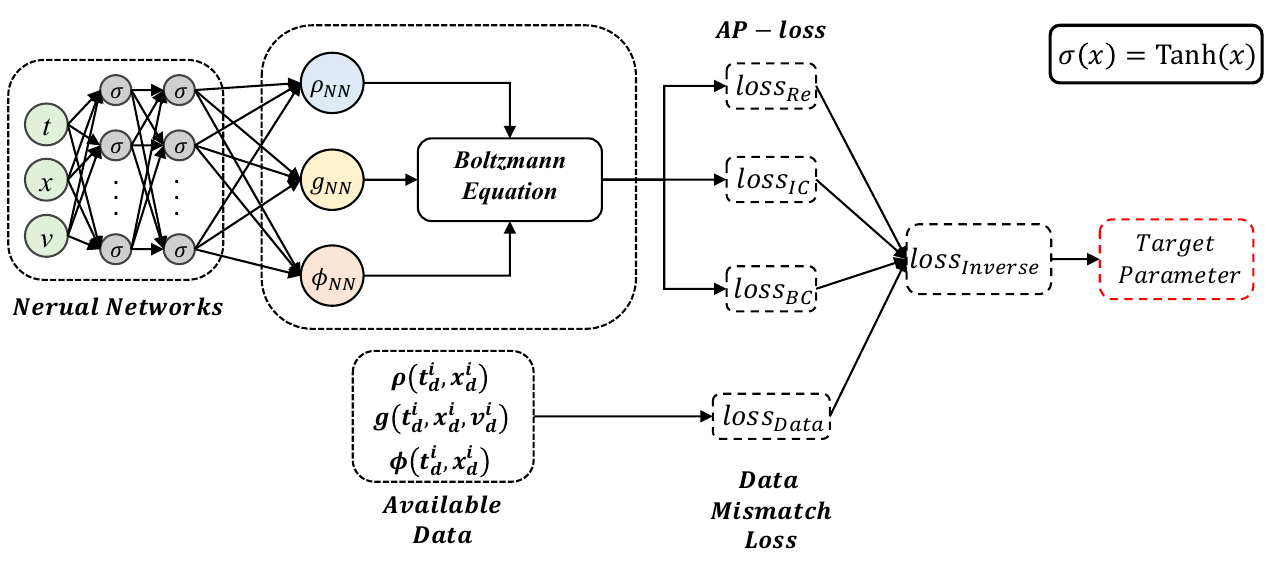}}
    \caption{APNN framework for solving the inverse problem under semi-conductor Boltzmann-(Poisson) equation. }
    \label{fig:inverse_framework}
\end{figure}

Similarly, Figure \ref{fig:inverse_framework} shows the framework of our method for inverse problems for the semi-conductor Boltzmann \eqref{Boltz-eqn} and the Boltzmann-Poisson system \eqref{eqn:BP} we studied in Section \ref{sec:Num}. 

\bigskip

Figure \ref{fig:0} below shows the work-flow of the APNNs method for the semi-conductor Boltzmann \eqref{Boltz-eqn} and Boltzmann-Poisson system \eqref{eqn:BP}. In the latter case, we need to construct an additional network to learn the potential function $\phi(t,x)$, which is temporal and spatial dependent. This is different from the semi-conductor Boltzmann model \eqref{Boltz-eqn}, where the external potential is given and fixed. 
We mention that the way we deal with $\phi$ for different model equations is applied to the inverse problem as well.

\begin{figure}[H]
\centerline{\includegraphics[width=1\linewidth]{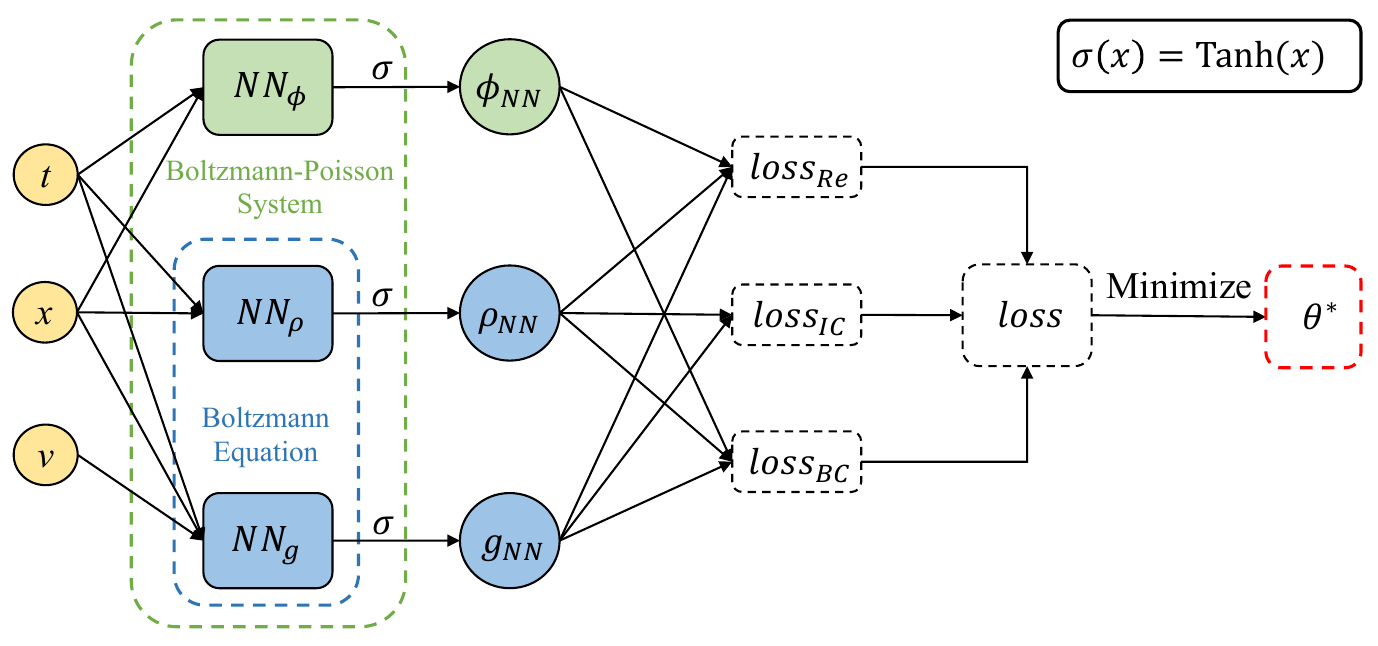}}
    \caption{APNNs work-flow for solving the semi-conductor Boltzmann-(Poisson) equation, with a difference in $\phi$ for different models. }
    \label{fig:0}
\end{figure}

%--------------------------------------

\section{Analysis results}
\label{sec:analysis}

\subsection{Preliminaries}

We first review an important theorem on the existence of the approximated neural network solution, namely the Universal Approximation Theorem (UAT) \cite{UAT} which uses a two-level neural network. We adapt it to discuss our model \eqref{Boltz-eqn} in the framework of APNN as shown in Figure \ref{fig:APNN}, and use the notations introduced in subsection \ref{sec:NN}. 

\begin{lemma}
\label{Lemma1}
Suppose the solution to \eqref{Boltz-eqn} satisfies $f \in C^1([0,T]) \cap C^1(\mathcal{D}) \cap C^1(\Omega)$. Let the activation function $\bar\sigma$ be any non-polynomial function in $C^1(\mathbb R)$, then for any $\delta>0$, there exists a two-layer neural network
$$ f^{NN}(t,x,v) = \sum_{i=1}^{m_1} w_{1i}^{(2)} \bar\sigma\left( \left(w_{i 1}^{(1)}, w_{i 2}^{(1)}, w_{i 3}^{(1)}\right) \cdot(t, x, v)+b_{i}^{(1)}\right) + b_{1}^{(2)}, 
$$
such that 
\begin{equation}
\label{UAT}
\begin{aligned}
\displaystyle &  \left\| f - f^{NN} \right\|_{L^{\infty}(K)} < \delta, \qquad 
\left\| \partial_t ( f - f^{NN}) \right\|_{L^{\infty}(K)} < \delta, \\[4pt]
& \left\| \nabla_x ( f - f^{NN}) \right\|_{L^{\infty}(K)} < \delta, \quad
\left\| \nabla_v ( f - f^{NN}) \right\|_{L^{\infty}(K)} < \delta, 
\end{aligned}
\end{equation}
where the domain $K$ denotes $[0,T]\times\mathcal{D}\times\Omega$. 
\end{lemma}
We remark that the above result can be generalized to neural network with several hidden layers \cite{Multilayer}. 

%-----------------------------------------------

\subsection{Convergence of the loss function}

Recall our APNN framework introduced in section \ref{sec:APNN}, we build the loss function based on the residual of the macro-micro system \eqref{Macro}--\eqref{Micro}, thus our neural network approximated solutions are $\rho$ and $g$, instead of $f$ if applying PINN to the model \eqref{Boltz-eqn}. 
We first show that a sequence of neural network solutions to 
\eqref{Macro}--\eqref{Micro} exists such that the total loss function converges to zero, if assuming the analytic solution 
\begin{equation}\label{condition}
\rho \in C^1([0,T])\cap C^1(\mathcal{D}), \qquad 
g \in C^1([0,T])\cap C^1(\mathcal{D}) \cap C^1(\Omega). 
\end{equation}

\begin{theorem}
\label{Thm1}
Consider the semiconductor Boltzmann equation \eqref{Boltz-eqn} with the given electric potential $\phi(t,x)$ and satisfies 
$ ||\nabla_x\phi||_{L^{\infty}([0,T]\times\mathcal{D})} \leq C_{\phi}$. Let $\rho$, $g$ be the analytic solution to the macro-micro system \eqref{Macro}--\eqref{Micro}, which are sufficiently smooth in its physical domain,
i.e., they satisfy the condition \eqref{condition}. 
Then there exists a sequence of neural network parameters $\{m_{[j]}, w_{[j]}, b_{[j]}\}_{j=1}^{\infty}$ such that the sequence of APNN solutions with $m_{[j]}$ nodes, denoted by 
$\{\rho_j(t,x) = \rho^{nn}(t,x; m_{[j]}, w_{[j]}, b_{[j]})\}_{j=1}^{\infty}$ and $\{g_j(t,x,v) = g^{nn}(t,x, v;m_{[j]}, w_{[j]}, b_{[j]})\}_{j=1}^{\infty}$ satisfy that 
$$ \mathcal{R}_{\mathrm{APNN}}^{\e}(\rho_j, g_j)\to 0, \quad \text{as  } j\to\infty. $$
\end{theorem}

{\bf Proof.}  
Define a sequence of small numbers $\delta_j = \frac{1}{j}$. 
By the UAT above, for any $\delta_j$, there exists a DNN solution $\rho_j(t,x)$, $g_j(t,x,v)$ 
such that 
$$ ||\rho_j - \rho||_{L^{\infty}(K_1)} < \delta_j, \qquad
||g_j - g||_{L^{\infty}(K_2)} < \delta_j, $$ 
and similarly for their first-order derivatives in $t, x$ for $\rho_j$ and $t,x,v$ for $g_j$. Here $K_1 = [0,T]\times \mathcal{D}, \, K_2 = [0,T]\times \mathcal{D} \times \Omega $. For simplicity, we denote the volume of $x$ and $v$ spaces by $\mathcal{D}$ and $\Omega$ below. Define 
$$ d_{ge,j}^{(1)} (t,x) = \partial_t \rho_j + 
\nabla_x \cdot \langle v g_j \rangle + \nabla_x \phi\cdot
\langle \nabla_v g_j \rangle. $$
Since 
$\partial_t \rho + 
\nabla_x \cdot \langle v g \rangle + \nabla_x \phi\cdot
\langle \nabla_v g \rangle = 0, $
subtract $d_{ge,j}^{(1)}$ by the above equation, then 
\begin{equation}
\label{d_GE1}
d_{ge,j}^{(1)}(t,x) = \partial_t (\rho_j - \rho) + 
\nabla_x \cdot \langle v (g_j-g) \rangle + \nabla_x \phi\cdot
\langle \nabla_v (g_j - g) \rangle. \end{equation}
Integrating $|d_{ge,j}^{(1)}|^2$ over $K_1$, one gets
\begin{equation}
\label{Loss1}
Loss^{(1)}_{\text{ge}} := || \partial_t (\rho_j - \rho) + 
\nabla_x \cdot \langle v (g_j-g) \rangle + \nabla_x \phi\cdot
\langle \nabla_v (g_j - g) \rangle ||_{L^2(K_1)}^2, 
\end{equation}
which corresponds to the first term in the loss function 
$\mathcal{R}_{\mathrm{APNN}}^{\e}$ shown in \eqref{Loss-APNN}.

Thanks to the boundedness of the physical domains, we have
\begin{equation*}
\begin{aligned}
\displaystyle
& ||\partial_t (\rho_j -\rho)||_{L^2(K_1)}^2  \leq ||\partial_t (\rho_j -\rho)||_{L^{\infty}(K_1)}^2 \cdot 
T \cdot \mathcal{D} \lesssim \delta_j^2, \\[6pt]
& ||\nabla_x \cdot \langle v (g_j-g) \rangle||_{L^2(K_1)}^2 
\leq ||\nabla_x \cdot \langle v (g_j-g) \rangle||_{L^{\infty}(K_1)}^2 \cdot C_{\Omega}\cdot T \cdot \mathcal{D} 
\lesssim \delta_j^2, \\[6pt]
& ||\nabla_x \phi\cdot\langle \nabla_v (g_j - g) \rangle||_{L^2(K_1)}^2 \leq 
C_{\phi}^2\, ||\nabla_v (g_j - g) \rangle||_{L^{\infty}(K_1)}^2 \cdot \Omega \cdot T \cdot \mathcal{D} \lesssim \delta_j^2,
\end{aligned}
\end{equation*}
where we used the fact that $L^2$ norm is bounded by $L^{\infty}$ norm, due to $(t,x,v)$ are bounded domains. 
Therefore, $Loss^{(1)}_{\text{ge}} \leq O(\frac{1}{j^2})$. 

\vspace{2mm}

Similarly, define  $$ d_{ge,j}^{(2)} = \e^2 \partial_t g_j + \e (I-\Pi)(v \cdot \nabla_x g_j + \nabla_x\phi \cdot \nabla_v g_j)
- 2 v \cdot \nabla_x\phi\, \rho_j M(v) + v\cdot \nabla_x \rho_j M(v) - Q(g_j). $$
Since $\rho$, $g$ solves the equation \eqref{Micro} which is linear, subtract $d_{ge,j}^{(2)}$ by \eqref{Micro}, we have
\begin{equation}
\label{d_GE2}
\begin{aligned}
\displaystyle
d_{ge,j}^{(2)} = & \e^2 \partial_t (g_j-g) + \e (I-\Pi)
\left\{ v \cdot \nabla_x (g_j-g) + \nabla_x\phi \cdot \nabla_v (g_j-g) \right\} \\[6pt]
& - 2 v \cdot \nabla_x\phi \, (\rho_j - \rho) M(v) + v\cdot \nabla_x (\rho_j - \rho) M(v) - Q(g_j) + Q(g). 
\end{aligned}
\end{equation}
Integrating $|d_{ge,j}^{(2)}|^2$ over $K_2$, one gets
\begin{equation}
\label{Loss2}
\begin{aligned}
\displaystyle
Loss^{(2)}_{\text{ge}} & := ||\e^2 \partial_t (g_j-g) + \e (I-\Pi)
\left\{ v \cdot \nabla_x (g_j-g) + \nabla_x\phi \cdot \nabla_v (g_j-g) \right\} \\[6pt]
& \quad - 2 v \cdot \nabla_x\phi \, (\rho_j - \rho) M(v) + v\cdot \nabla_x (\rho_j - \rho) M(v) - Q(g_j) + Q(g)||_{L^2(K_2)}^2. 
\end{aligned}
\end{equation}
For the first four terms, we have
\begin{equation*}
\begin{aligned}
\displaystyle
& \quad \e^4 ||\partial_t (g_j-g)||_{L^2(K_2)}^2 \leq 
\e^2 ||\partial_t (g_j-g)||_{L^{\infty}(K_2)}^2 \cdot \Omega \cdot T \cdot \mathcal{D} \lesssim \e^4 \delta_j^2, \\[6pt]
& \quad \e^2 || (I - \Pi) \left\{ v \cdot \nabla_x (g_j-g) + \nabla_x\phi \cdot \nabla_v (g_j-g) \right\} ||_{L^2(K_2)}^2 \leq \e^2 || v \cdot \nabla_x (g_j-g) + \nabla_x\phi \cdot \nabla_v (g_j-g) ||_{L^2(K_2)}^2   \\[6pt]
& \qquad \leq \e^2 \left( ||\nabla_x (g_j-g)||_{L^{\infty}(K_2)}^2 \cdot C_{\Omega} \cdot T \cdot \mathcal{D} + C_{\phi}^2 \, 
||\nabla_v (g_j-g)||_{L^{\infty}(K_2)}^2 \cdot \Omega \cdot T \cdot \mathcal{D} \right) \lesssim \e^2 \delta_j^2, \\[6pt]
& \quad || 2 v \cdot \nabla_x\phi \, (\rho_j - \rho) M(v) ||_{L^2(K_2)}^2 \leq 4 C_{\phi}^2 \, ||\rho_j - \rho||_{L^{\infty}(K_2)}^2
\cdot \tilde{C}_{\Omega} \cdot T \cdot \mathcal{D} \lesssim \delta_j^2, \\[6pt]
& \quad || v\cdot \nabla_x (\rho_j - \rho) M(v) ||_{L^2(K_2)}^2 \leq
|| \nabla_x (\rho_j - \rho) ||_{L^{\infty}(K_2)}^2 \cdot \tilde{C}_{\Omega} \cdot T \cdot \mathcal{D} \lesssim \delta_j^2. 
\end{aligned}
\end{equation*}
For the last two terms in \eqref{Loss2}, thanks to the bounded and linear operator $Q$, one can derive 
$$ ||Q(g_j) - Q(g)||_{L^2(K_2)}^2 = || Q(g_j-g)||_{L^2(K_2)}^2 
\lesssim ||g_j - g||_{L^2(K_2)}^2 \leq || g_j - g||_{L^{\infty}(K_2)}^2 \cdot \Omega \cdot T \cdot \mathcal{D} \lesssim \delta_j^2. 
$$
To sum up the above terms, note that $\delta_j = \frac{1}{j}$, we have
$$ Loss^{(2)}_{\text{ge}} \lesssim \max\{1, \e^2, \e^4\} \delta_j^2 \leq O(\delta_j^2).  $$
Thus the loss arisen from the governing equations is bounded by $O(\delta_j^2)$, namely
\begin{equation}\label{Loss_GE} 
Loss^{(1)}_{\text{ge}} + Loss^{(2)}_{\text{ge}} \leq O(\delta_j^2) = O(\frac{1}{j^2}).  
\end{equation}
For the inflow boundary data, denote $f_j = \rho_j M + \e g_j$, 
then $Loss_{\text{BC}}$ is bounded by
\begin{equation}\label{Loss_BC}
Loss_{\text{BC}}:= || f_j - f||_{L^2(\gamma_{T}^{-})}^2 \leq 
|| f_j - f||_{L^{\infty}(\gamma_{T}^{-})}^2 \cdot \Omega \cdot T 
\cdot |\partial\mathcal{D}| \lesssim \delta_j^2, 
\end{equation}
where $\gamma_{-}$ is defined in \eqref{gamma} and $\gamma_{T}^{-} = [0,T]\times \gamma_{-}$. For the case of specular
reflection BC, the proof is similar and omitted here. 

For the initial data, let the APNN approximation be 
$f_j(0,x,v) = \rho_j(0,x) M(v) + \e g_j(0,x,v)$, then
\begin{equation}\label{Loss_IC}
\begin{aligned}
\displaystyle
Loss_{\text{IC}} & := || f_j(0,x,v) - f_{\text{IC}}(x,v)||_{L^2(\mathcal{D}\times\Omega)^2}^2 \\[6pt]
& \leq 
|| f_j(0,x,v) - f_{\text{IC}}(x,v)||_{L^{\infty}(\mathcal{D}\times\Omega)^2}^2 \cdot \Omega \cdot \mathcal{D}
\lesssim \delta_j^2. 
\end{aligned}
\end{equation}
Combine \eqref{Loss_GE}, \eqref{Loss_BC} and \eqref{Loss_IC}, we conclude that the total loss function for the APNN method 
$$ \mathcal{R}_{\mathrm{APNN}}^{\e} = Loss_{\text{ge}}^{(1)} + Loss_{\text{ge}}^{(2)} + Loss_{\text{BC}} + Loss_{\text{IC}} 
\leq O(\frac{1}{j^2}). $$
Therefore, $\mathcal{R}_{\mathrm{APNN}}^{\e}(\rho_j, g_j) \to 0$ as $j\to \infty$. 

%----------------------------------------------
\subsection{Convergence of the APNN solution}

Now we prove that with the parameters $\{m_{[j]}, w_{[j]}, b_{[j]}\}_{j=1}^{\infty}$ equipped, the asymptotic-preserving neural network in Theorem \ref{Thm1} indeed converges to the analytic solution of the semiconductor Boltzmann model \eqref{Boltz-eqn}. 

\begin{theorem}
\label{Thm2}
Let $\{m_{[j]}, w_{[j]}, b_{[j]}\}_{j=1}^{\infty}$ be a sequence given in Theorem \ref{Thm1}, and $f$ be the analytic solution to the linear Boltzmann equation \eqref{Boltz-eqn}.  Denote $f_j = \rho_j M + \e g_j$, with the sequence of APNN solutions $\{\rho_j\}_{j=1}^{\infty}$, $\{g_j\}_{j=1}^{\infty}$ defined in Theorem \ref{Thm1}. 
Then, $\mathcal{R}_{\mathrm{APNN}}^{\e}(\rho_j, g_j)\to 0$ implies that for any $t\in (0,T]$, 
\begin{equation}\label{eqn:thm2}
  ||f_j(t, \cdot, \cdot, m_{[j]}, w_{[j]}, b_{[j]}) - f(t,\cdot,\cdot) ||_{H^1(\mathcal{D}\times \Omega)}\leq \frac{C}{j^2}\, e^{-K t}\,.
\end{equation}  
Moreover, we conclude that
$$ ||f_j(\cdot, \cdot, \cdot, m_{[j]}, w_{[j]}, b_{[j]}) - f ||_{L^{\infty}\left([0,T]; H^1(\mathcal{D}\times \Omega)\right)}
\to 0, \, \text{  as  } j\to\infty. $$
\end{theorem}

{\bf Proof. }
Multiply \eqref{d_GE1} by $M(v)$ and 
divide \eqref{d_GE2} by $\e$ on both sides, then add them up we have
\begin{equation}
\label{eqn:fj}
\begin{aligned}
\displaystyle
\partial_t (f_j - f) + & \frac{1}{\e}\left\{ v\cdot \nabla_x (f_j-f) + \nabla_x\phi \cdot
\nabla_v (f_j - f)\right\} \\[6pt]
& = d_{ge,j}^{(1)}M - \frac{1}{\e}d_{ge,j}^{(2)} - \frac{1}{\e^2}Q(f_j - f), 
\end{aligned}
\end{equation}
where $f = \rho M + \e g$ and $f_j = \rho_j M + \e g_j$ are used. 

Inspired by \cite[section 3.2]{Mouhot06} where the model without scaling is discussed, we consider the following fluctuations around equilibrium for distribution functions $f$ and $f_j$, 
\begin{equation}\label{fluc:f}
f = f_{\infty} + \e \sqrt{f_{\infty}}\, h, \qquad 
f_j = f_{\infty} + \e \sqrt{f_{\infty}}\, h_j, 
\end{equation}
where the stationary solution is given by 
$$ f_{\infty} = e^{\phi}\frac{\rho_{\infty}}{|| e^{\phi} ||_{L^1_x}}\mathcal{M}, $$
with $\rho_{\infty}$ the total mass of $f$ that is conserved as the initial data, and $\mathcal{M}$ is the normalized Maxwellian
$\mathcal{M}(v) = \frac{e^{-|v|^2/2}}{(2 \pi)^{d/2}}$. 
In addition, we assume that the potential function $\phi$ satisfies 
$\displaystyle ||\phi ||_{C^2(\mathcal{D})}\leq C_{\phi}$. 

\vspace{2mm}
Inserting \eqref{fluc:f} into \eqref{eqn:fj}, one gets the linearized equation for 
$\tilde h:= h - h_j$ as follows: 
\begin{equation} 
\label{eqn:hj}
\partial_t \tilde h + \frac{1}{\e}\left( v\cdot \nabla_x \tilde h + \nabla_x\phi \cdot
\nabla_v \tilde h \right) = \underbrace{ -d_{ge,j}^{(1)}M + \frac{1}{\e}d_{ge,j}^{(2)}}_{\text{Term A}} + \frac{1}{\e^2} Q(\tilde h). 
\end{equation}
Define the operator $$ G_{\e}:=\frac{1}{\e^2}Q - \frac{1}{\e}v\cdot\nabla_x, $$ and we take $\tilde{h}_{\text{IC}} \in 
H_{x,v}^1 \cap \text{Ker}(G_{\e})^{\perp}$. 
For simplicity, in the following energy estimates we only consider the periodic BC. Studying the cases of incoming and specular reflection BC are challenging and will be considered a future work. 

Multiply \eqref{eqn:hj} by $\tilde h$ and integrate it on $\mathcal{D}\times\Omega$, then
\begin{equation}
\label{eqn2:hj}
\frac{1}{2}\frac{d}{dt} ||\tilde h||_{L^2}^2 
 = \langle \text{Term A}, \tilde h \rangle_{L^2} + \frac{1}{\e^2}  \langle Q(\tilde h),\, \tilde h  \rangle_{L^2},  
\end{equation}
where we assumed vanishing boundary condition in velocity for $\tilde h$. 

Thus, for the time evolution for the $L^2$ norm of $\tilde h$, we have
\begin{equation*}
\frac{d}{dt}||\tilde h||_{L^2}^2 \leq - \frac{2\lambda}{\e^2}
||\tilde h - \Pi\tilde h ||_{\Lambda}^2 + \frac{C_1}{\e}
\left( \e ||d_{\text{ge},j}||_{L^2}^2 + \frac{1}{\e} ||\tilde h||_{L^2}^2 \right),  \label{h:l2} 
\end{equation*}
where the Young's inequality is used in the last term. Similarly, the time evolution for the $L^2$ norm of $\nabla_x \tilde h$ is given by
\begin{equation*} \frac{d}{dt}||\nabla_x \tilde h ||_{L^2}^2 \leq
-\frac{2\lambda}{\e^2} ||\nabla_x \tilde h^{\perp}||_{\Lambda}^2 + \frac{2 C_{\phi}}{\e} ||\nabla_v \tilde h||_{L^2}\cdot 
|| \nabla_x \tilde h||_{L^2}\,.  \label{hx}
\end{equation*}
For the time evolution for the $L^2$ norm of $\tilde h^{\perp}:= \tilde h - \Pi \tilde h$, one has
\begin{equation}
\label{h:perp}
\begin{aligned}
 \frac{d}{dt} ||\nabla_v \tilde h^{\perp}||_{L^2}^2 & \leq  \frac{C}{\e^2} ||\tilde{h}^{\perp}||_{\Lambda}^2 
- \frac{C^{'}}{\e^2}  ||\nabla_v \tilde{h}^{\perp}||_{\Lambda}^2 +  C^{''} ||\nabla_x \tilde h ||_{L^2}^2 
 + \frac{\tilde C^{'}}{\e}\left(  \e ||d_{\text{ge},j}||_{L^2}^2 + \frac{1}{\e} ||\nabla_v \tilde h^{\perp}||_{L^2}^2 \right) \\[4pt]
& + \frac{1}{\e}\left( ||\nabla_v\tilde h||_{L^2}^2 + ||\tilde h||_{L^2} \cdot ||\nabla_v\tilde{h}^{\perp}||_{L^2} + ||\tilde h||_{L^2}\cdot ||\nabla_v\tilde h||_{L^2} \right). 
\end{aligned}
\end{equation}
where some properties of $\nabla_v Q$ are used, see \cite[section 1.2 and Appendix B.2.6]{Mouhot06} for details. 

For the time evolution of the mixed term, one has
\begin{equation}
\label{h:mixed}
\frac{d}{dt}\langle \nabla_x\tilde h, \nabla_v\tilde h \rangle_{L^2}
 \leq \frac{C}{\e^3}||\nabla_x \tilde{h}^{\perp}||_{\Lambda}^2 
+ \frac{C^{'}}{\e}||\nabla_v\tilde{h}^{\perp}||_{\Lambda}^2  - \frac{1}{2\e} ||\nabla_x\tilde h||_{L^2}^2 
+ \frac{\tilde C_{\phi}}{\e}||\nabla_v\tilde h||_{L^2}^2\,, 
\end{equation}
where we used properties of $Q$ and Young's inequality as follows
\begin{equation*}
\begin{aligned}
\displaystyle
& \quad \langle Q(\nabla_x \tilde h), \nabla_v\tilde h\rangle  = 
\langle Q(\nabla_x \tilde h - \Pi(\nabla_x \tilde h)), \nabla_v\tilde h\rangle \\[6pt]
&  \leq C ||\nabla_x \tilde h^{\perp} ||_{\Lambda}\cdot 
|| \nabla_v\tilde h ||_{\Lambda} 
\leq  C \eta ||\nabla_x \tilde h^{\perp} ||_{\Lambda}^2 + \frac{C}{\eta} ||\nabla_v \tilde h||_{\Lambda}^2, 
\end{aligned}
\end{equation*}
together with the relation 
$$ ||\nabla_v\tilde h||_{\Lambda}^2 
\leq 2 ||\nabla_v \tilde{h}^{\perp}||_{\Lambda}^2 
+ 2 ||\nabla_v \Pi(\tilde h)||_{\Lambda}^2 \leq 
2 ||\nabla_v \tilde{h}^{\perp}||_{\Lambda}^2 + 2 C_L ||\nabla_x h||_{L^2}^2. $$ 
See \cite[Appendix B.2.7]{Briant} for details. Note that different from \cite{Briant}, we have the source term on the RHS and the potential in the convection on the LHS of equation \eqref{eqn:hj}. 

As suggested by \cite{Briant}, we define the norm
\begin{equation}
 ||\tilde h||_{\mathcal{H}_{\e^{\perp}}^1}^2 := 
 a_1 ||\tilde h||_{L^2}^2 + a_2 ||\nabla_x \tilde h||_{L^2}^2 
 + a_3 ||\nabla_v \tilde h^{\perp}||_{L^2}^2 + a_4 \,\e \langle \nabla_x \tilde h, \nabla_v \tilde h \rangle_{L^2}, 
\end{equation}
with $a_1$, $a_2$, $a_3$ and $a_4$ strictly positive and 
$ ||\cdot||_{\mathcal{H}_{\e^{\perp}}^1} \sim ||\cdot||_{H^1_{x,v}}$. 
Here all the positive constants $C$, $C^{'}$, $C^{''}$ are generic. 
For our linear semiconductor model \eqref{Boltz-eqn}, the collision operator $Q$ satisfies all the assumptions in \cite[section 1.2]{Mouhot06} by taking $||\cdot ||_{\Lambda}= ||\cdot||_{L^2_{x,v}}$. 

Then adding up \eqref{h:l2}, \eqref{hx}, \eqref{h:perp} and \eqref{h:mixed} multiplied by $\e$, one has
\begin{equation}
\begin{aligned}
\displaystyle
 \frac{d}{dt} ||\tilde h||_{\mathcal{H}_{\e^{\perp}}^1}^2 
 & \leq 
 - \frac{K_1}{\e^2}\left( ||\tilde h^{\perp}||_{\Lambda}^2 + 
 ||\nabla_x\tilde{h}^{\perp}||_{\Lambda}^2 + ||\nabla_v \tilde{h}^{\perp}||_{\Lambda}^2 \right)- K_2 ||\nabla_x \tilde h||_{L^2}^2 
 + K_3 ||d_{\text{ge},j}||_{L^2}^2  \\[6pt]
 & \quad + \frac{C_{\phi}^{'}}{\e^2}\left( ||\tilde h||_{L^2}^2 + ||\nabla_v \tilde{h}^{\perp}||_{\Lambda}^2 + 
  ||\nabla_x\tilde h||_{L^2}^2 \right)
 + C_{\phi}^{''} ||\nabla_v \tilde{h}^{\perp}||_{L^2}^2 \\[6pt]
 &  \leq - K_0 ||\tilde h||_{H_{\Lambda}^1}^2 + K_3 ||d_{\text{ge},j}||_{L^2}^2 \leq - K_0 ||\tilde h||_{H^1}^2 + K_3 ||d_{\text{ge},j}||_{L^2}^2 \,, 
\end{aligned}
\end{equation}
under the reasonable assumption that $||\phi||_{C^2(\mathcal{D})}$ is bounded and small. The Poincaré inequality is used and we refer to \cite[Section 7.2]{Briant} for details on the second before last inequality. 
 
Therefore by the Gr\"onwall’s inequality and 
$ \int_0^t ||d_{\text{ge},j}(\tau,\cdot,\cdot,\cdot)||_{L^2}^2\, d\tau = Loss_{\text{GE}}$, then 
\begin{equation}
\begin{aligned} 
\displaystyle
 || f_j(t,\cdot,\cdot) - f(t,\cdot,\cdot) ||_{H^1(\mathcal{D}\times\Omega)}^2 
& \leq C\, e^{-K t} 
\left\{ Loss_{IC} + Loss_{\text{GE}} + Loss_{\text{BC}}\right\} 
\\[6pt]  
& \leq C\, e^{-K t}\,\mathcal{R}_{\mathrm{APNN}}^{\e}
\leq  \frac{C}{j^2}\, e^{-K t}\,,
\end{aligned}
\end{equation}
for any $t \in (0,T]$ and positive constants $C$, $K$ that are independent of $\e$. Notice that this convergence owns an {\it exponentially} decay in time. 

Taking $L^{\infty}$ norm in $t \in [0, T]$, we conclude that
$$ || f_j - f ||_{L^{\infty}\left([0,T]; H^1(\mathcal{D}\times\Omega)\right)} \to 0, \qquad \text{as  } j\to \infty. $$
This indicates that when the total loss function  $\mathcal{R}_{\mathrm{APNN}}^{\e}(\rho_j, g_j)\to 0$ as $j\to\infty$, the sequence of APNN solutions $\rho_j$, $g_j$ leading to $f_j$ also converges to the true solution $f$ of the semi-conductor Boltzmann model \eqref{Boltz-eqn}. Our convergence result is {\it uniform} in $\e$, and can be extended to higher Sobolev space $H^k_{x,v}$ by using the strategy in \cite{Briant}. 

We also remark that one could adapt our analysis in Section \ref{sec:analysis} similarly to other kinetic models, such as linear transport model, the Fokker-Planck equation and the nonlinear Boltzmann or Landau equations, thanks to \cite{Mouhot06} where the properties of collision operators in those models were carefully studied. In our future work, we will further improve our analysis result by studying the Barron-type functions \cite{Gu-Ng} and posterior estimates of the solution error.

%---------------------------------------------------------------
\section{Numerical Examples}
\label{sec:Num}
In our numerical experiments, we will show several examples to illustrate effectiveness of our designed APNN method. For the first two numerical examples, we employ the APNN for the deterministic semiconductor Boltzmann model \eqref{Boltz-eqn} and the Boltzmann-Poisson system \eqref{eqn:BP}. We compare the performance of PINN and APNN methods in different regimes, ranging from the kinetic regime ($\e\approx O(1)$) to the diffusive regime ($\e \ll 1$), and observe that the APNN can capture the multiscale nature of the model thanks to the design of the loss function based on the macro-micro decomposition.

\textbf{Networks Architecture.} In our experiments, we approximate the solution by  the feed-forward neural network (FNN) with one input layer, one output layer and $4$ hidden layers with $128$ nerouns in each layer, unless otherwise specified. The hyperbolic tangent function (Tanh) is chosen as our activation function. 

\textbf{Training Settings.} We train the neural network by Adam with Xavier initialization. We set epochs to be $10000$ and the learning rate to be $10^{-3}$, and use full batch for most of the following experiments in the numerical experiments unless otherwise specified. All the hyper-parameters are chosen by trial and error.  

\textbf{Loss Design.} We show the losses for PINNs and APNNs in  \eqref{empirical_Loss-PINN} and \eqref{empirical_Loss-APNN}. For most of our experiments, we consider the spatial and temporal domains to be $[0,1]$ and $[0,0.1]$ respectively.  We choose the collocation points $\{(t_i, x_i, v_i)\}$ for $f(t,x,v)$ in the following way. For spatial points $x_i$, we select $99$ interior points evenly spaced in $[0, 1]$.  For temporal points $t_i$, we select $20$ interior points evenly spaced in the range $[0,0.1]$. We use the tensor product grid for the collocation points. For velocity points $v_i$, $N_v = 8$ points are generated by the Hermite quadrature rule. The penalty parameters in \eqref{Loss-APNN} are set to $(\lambda_1, \lambda_2) = (1,1)$. 

The reference solutions are obtained by the AP scheme \cite{JP2000} 
with $\Delta x=0.01$ and $\Delta t =5\times 10^{-5}$. For the velocity discretization, the integral in velocity for the second equation of \eqref{MM} is computed by the Hermite quadrature rule with $N_v=8$ quadrature points. We adopt the Hermite polynomials in velocity discretization for both APNN and traditional methods to provide more accurate velocity derivatives, a technique similar to the moment method \cite{JP00,SZ99}. See details in the Appendix.  

We compute the relative $\ell^2$ or $\ell^{\infty}$ error of the density $\rho(t,x)$ between our proposed neural network approximations and reference solutions, with the relative $\ell^2$ error defined by: 
\begin{equation}
\mathcal{E(t)}:=\sqrt{\frac{\sum_j|\rho_{\theta}^{\text{NN}}(t,x_j)-\rho^{\text{ref}}(t,x_j)|^2}{\sum_j|\rho^{\text{ref}}(t,x_j)|^2}}.
\end{equation}
We run our experiments on a server with Intel(R) Xeon(R) Gold $6230$ and GPU Quadro RTX $8000$.

%------------------------------------------------------
\subsection{Problem I: given the potential}
We first consider the following setup for the model equation \eqref{Boltz-eqn}:
$$ x\in [0,1], \qquad \phi = e^{-50 exp(1) (1/4-x)^2}, \qquad \sigma(v,w)=2. $$
Assume the initial data $\displaystyle f(t=0,x,v) = \frac{1}{\sqrt{\pi}} e^{-v^2}$ and incoming boundary conditions in space. \\

\begin{figure}[H]
    \centering
    \begin{minipage}[b]{0.49\textwidth}
        \centering
        \includegraphics[width=\textwidth]{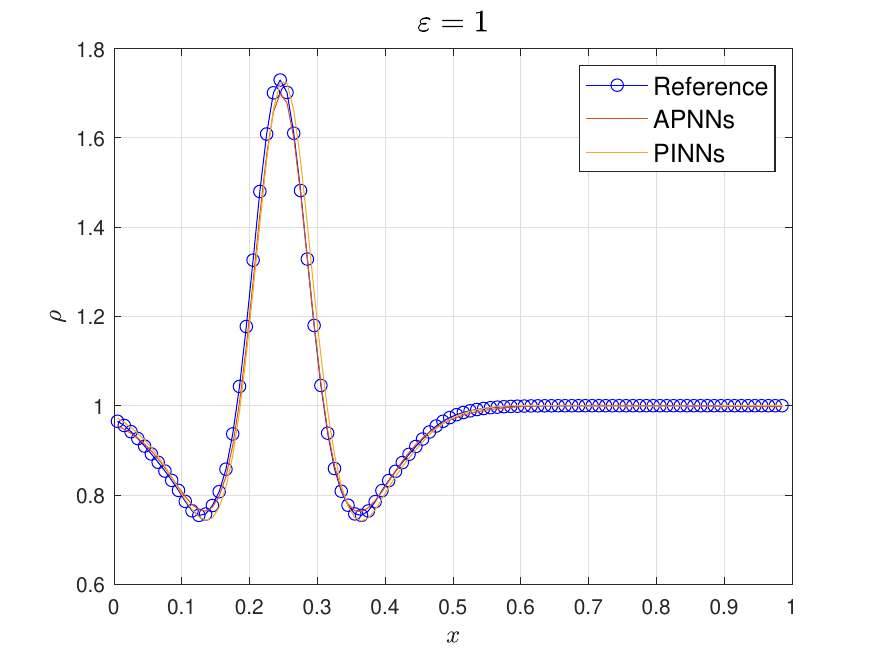} (a)
    \end{minipage}
    \begin{minipage}[b]{0.49\textwidth}
        \centering
        \includegraphics[width=\textwidth]{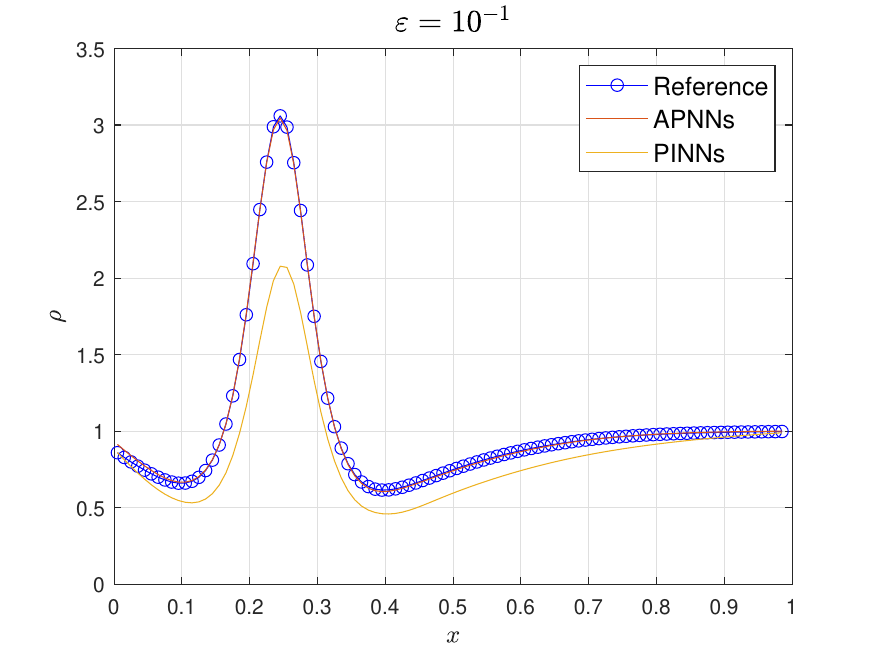} (b)
    \end{minipage}
    \begin{minipage}[b]{0.49\textwidth}
        \centering
        \includegraphics[width=\textwidth]{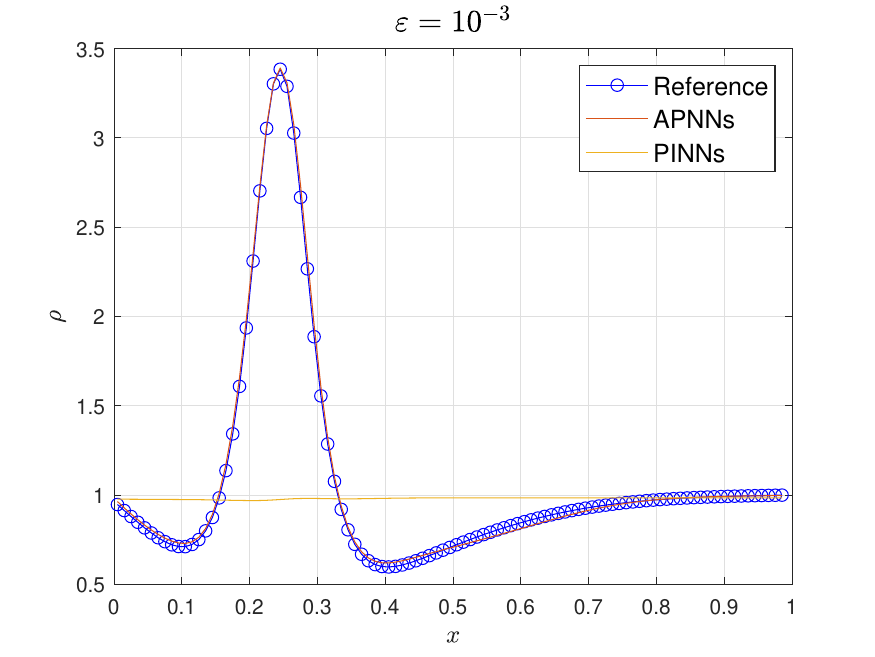} (c)
    \end{minipage}
    \begin{minipage}[b]{0.49\textwidth}
        \centering
        \includegraphics[width=\textwidth]{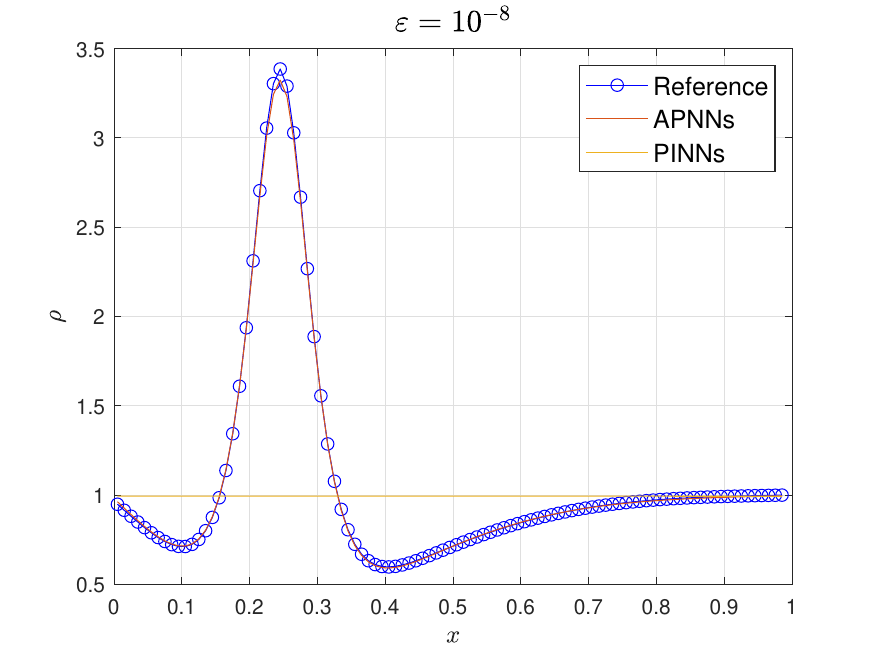} (d)
    \end{minipage}
\caption{Problems I with different $\e$. 
Density $\rho$ for PINNs, APNNs and reference solutions at $T=0.1$. }
    \label{fig:test1}
\end{figure}

\noindent\textbf{Forward Problem}. We first study the forward problems to compare PINNs and APNNs with different $\e$ in the model. In Figure \ref{fig:test1}, we plot the solutions of $\rho$ at the final time $T=0.1$, generated by PINNs or APNNs, and the reference solution computed by the traditional AP scheme \cite{JP2000}. We observe that when $\e$ is relatively large, e.g. $\e=1$, the performance of PINNs and APNNs agree with the reference solution. However, as $\e$ decreases, the performance of PINNs tends to be worse compared with APNNs, particularly when $\e \ll 1$, the approximated $\rho$ obtained by PINNs converges to a trivial solution, which is a constant in our problem setting. In constrast, APNNs can still capture an accurate solution compared to the reference one, thanks to the AP property of the loss function in APNNs. 
To qualitatively demonstrate the performance of the APNNs, we report the relative $\ell^2$ errors for the test sets
generated by PINNs and APNNs. It can be seen that the relative $\ell^2$ error of APNNs is much smaller than that of PINNs. 
\begin{table}[htbp]
\centering
\begin{tabular}{|p{1.5cm}|p{1.5cm}|p{1.5cm}|p{1.5cm}|p{1.5cm}|}
\hline
$\mathbf{\e}$ & $\mathbf{1}$ & $\mathbf{10^{-1}}$ &  $\mathbf{10^{-3}}$ & $\mathbf{10^{-8}}$\\
\hline
PINNs &\scriptsize $2.47 \times 10^{-2}$  &\scriptsize $2.52 \times 10^{-1}$  & \scriptsize$5.10 \times 10^{-1}$ &\scriptsize $5.05 \times 10^{-1}$\\
\hline
APNNs & \scriptsize $\mathbf{1.26 \times 10^{-2}}$  & \scriptsize $\mathbf{9.70 \times 10^{-3}}$  & \scriptsize $\mathbf{1.89 \times 10^{-2}}$ & \scriptsize $\mathbf{1.07 \times 10^{-2}}$ \\
\hline
\end{tabular}
\caption{Problems I. Relative $\ell^2$ error comparison for PINNs and APNNs with different $\e$ at the final time at $T=0.1$. }
\label{tab:test1}
\end{table}

\begin{figure}[H]
    \centering
    \includegraphics[width=0.49\textwidth]
{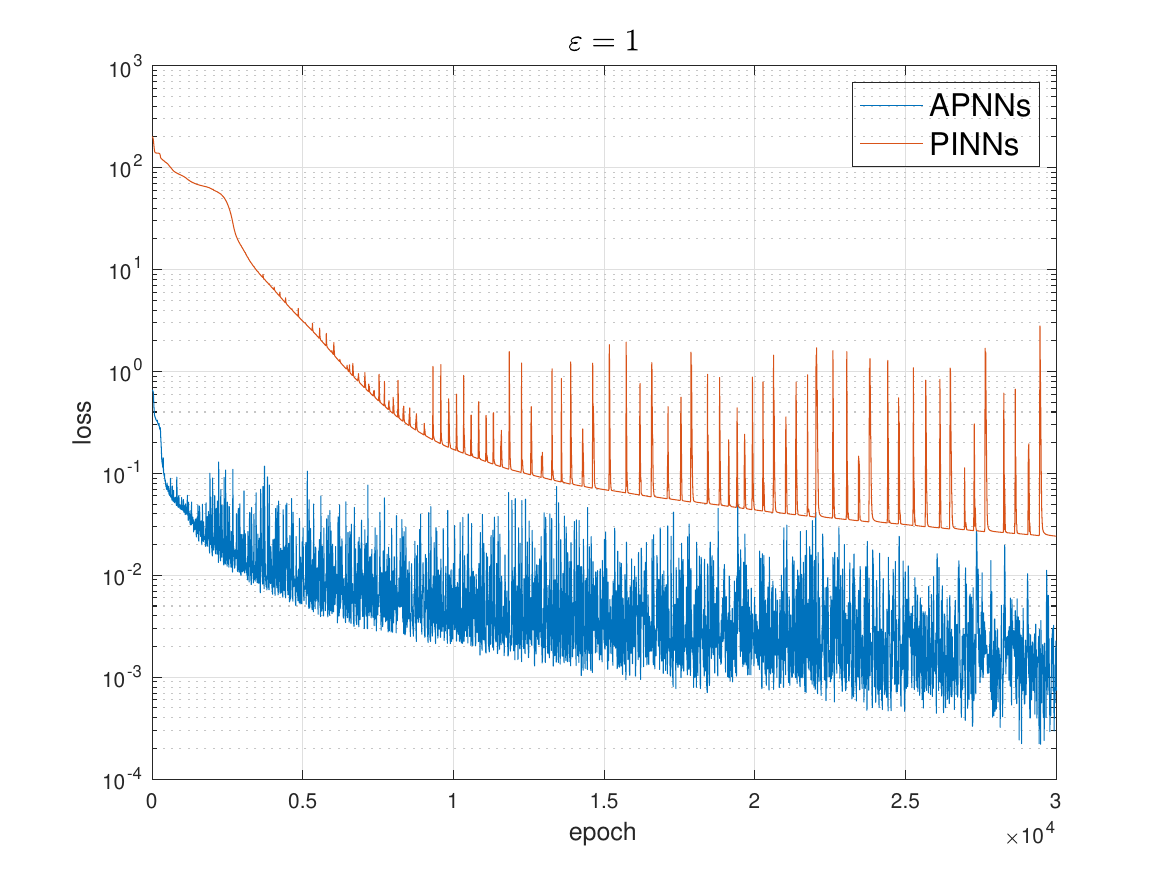}
    \includegraphics[width=0.49\textwidth]{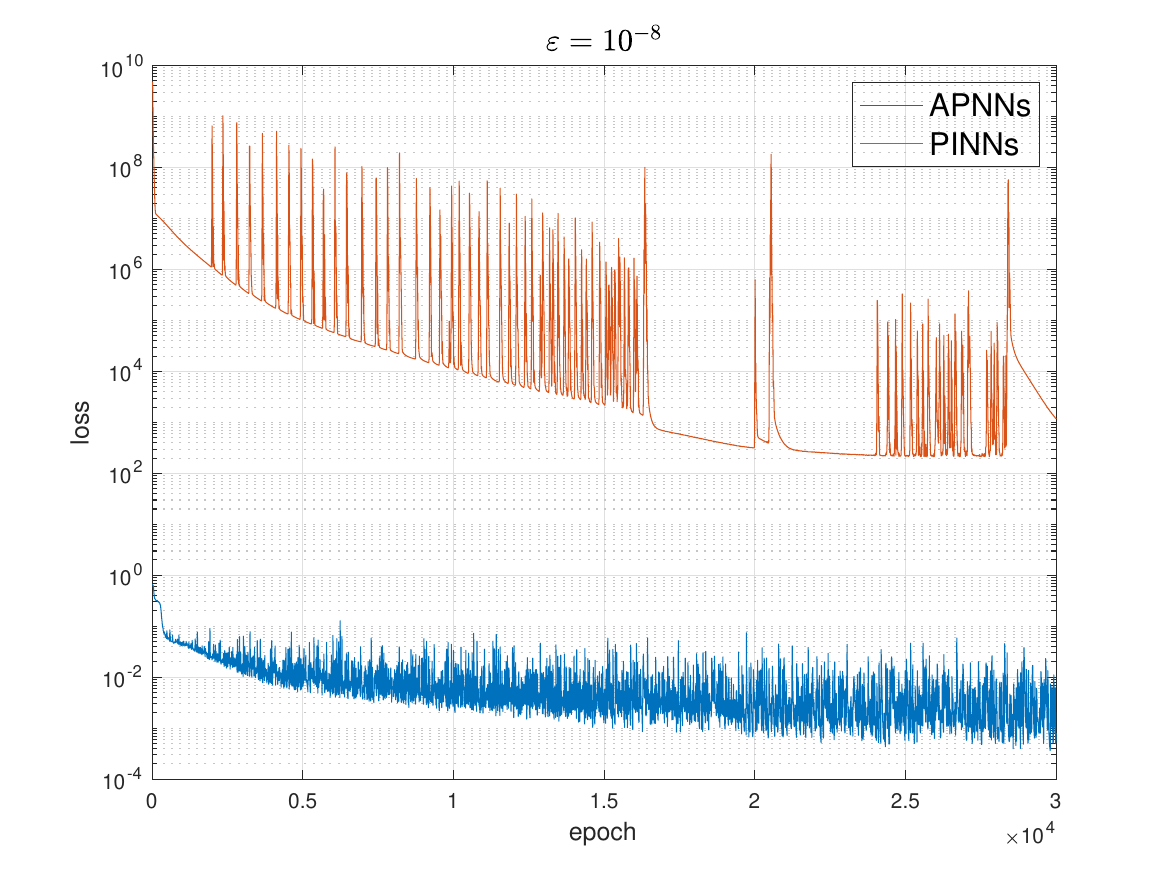}    
    \caption{Problem I with $\e=1
    % , 10^{-1}, 10^{-3}
    , 10^{-8}$. Comparison of training losses between PINNs and APNNs. }
    \label{fig:training loss}
\end{figure}

To further demonstrate that the APNNs method outperforms than PINNs especially when $\e$ is small, we plot the training loss in Figure \ref{fig:training loss}, with the number of epochs to $30000$.
It can be found that during the training, the  oscillations of the loss of PINNs are much larger than APNNs even when their training losses  converged, especially when $\e$ is small. Specifically, when $\e=1$, the  oscillation of the loss for both methods are similarly around two orders of magnitude. However, when $\e=10^{-8}$, while the  oscillations of PINNs's loss are around five-orders of magnitude, the  oscillations of APNNs's loss remian two orders of magnitude. This indicates that PINNs struggle to capture the multiscale nature of the underlying dynamics.
 
\noindent \textbf{Inverse Problem. } 
We now consider an inverse problem inferring the scattering coefficient from available measurement data using the APNNs formulation. We consider two possible scenarios of the observation data when employing APNNs for inverse problems: full and partial observation data. We will conduct a series of experiments to evaluate the performance of PINNs and APNNs in different regimes. For comparison, we also solve the inverse problem by applying the standard PINNs residual in the loss function as shown below. 

\bigskip

\noindent {\bf Scenario 1: Full observation data. } 
First, we consider an ideal inverse problem setup inferring the scattering coefficient $\sigma$ by assuming the availability of a limited number of the full observation dataset. We obtain synthetic dataset for $\rho(t_d^i,x_d^i)$ and $f(t_d^i,x_d^i,v_d^i)$ 
by traditional AP numerical scheme developed in \cite{JP2000} and compute $g(t_d^i,x_d^i,v_d^i)$ by using 
\begin{equation*}
g(t_d^i,x_d^i,v_d^i) = \dfrac{f(t_d^i,x_d^i,v_d^i)-\rho(t_d^i,x_d^i)}{\e}. 
\end{equation*}

For APNNs, we add a data misfit loss as below 
to the original empirical loss function \eqref{empirical_Loss-APNN}: 
\begin{equation}
    L_{d}^{APNN}(\theta) = \frac{\omega_d^\rho}{N_{d_1}}\sum_{i=1}^{N_{d_1}}\left|\rho^{NN}(t_d^i,x_d^i;\theta)-\rho(t_d^i,x_d^i)\right|^2+\frac{\omega_d^g}{N_{d_2}}\sum_{i=1}^{N_{d_2}}\left|g^{NN}(t_d^i,x_d^i,v_d^i;\theta)-g(t_d^i,x_d^i,v_d^i)\right|^2. 
    \label{Inverse_APNN}
\end{equation}
For PINNs, we add a data misfit loss as below to the original empirical loss function \eqref{empirical_Loss-PINN}:
\begin{equation}
    L_{d}^{PINN}(\theta) = \frac{\omega_d^f}{N_{d_3}}\sum_{i=1}^{N_{d_3}}\left|f^{NN}(t_d^i,x_d^i,v_d^i;\theta)-f(t_d^i,x_d^i,v_d^i)\right|^2
    \label{Inverse_PINN}
\end{equation}
Where $\omega_d^\rho, \omega_d^g, \omega_d^f$ 
are penalty parameters and set to be $1$ for each empirical risk loss term
in \eqref{Inverse_APNN} and \eqref{Inverse_PINN}. 
We train the network model on measurements with $N_{d_1} = 100$ samples randomly selected in the domain $(t,x) \in [0,0.1] \times [0,1]$, and $N_{d_2}, N_{d_3} = 100$ are samples randomly selected in the domain $(t,x,v) \in [0,0.1] \times [0,1] \times \Omega_v$, where $\Omega_v$ is the discrete points in the velocity space obtained by the Hermite quadrature rule from which $20 \%$ points are randomly selected for validation.

\begin{figure}[H]
    \centering
    \includegraphics[width=0.49\textwidth]
    {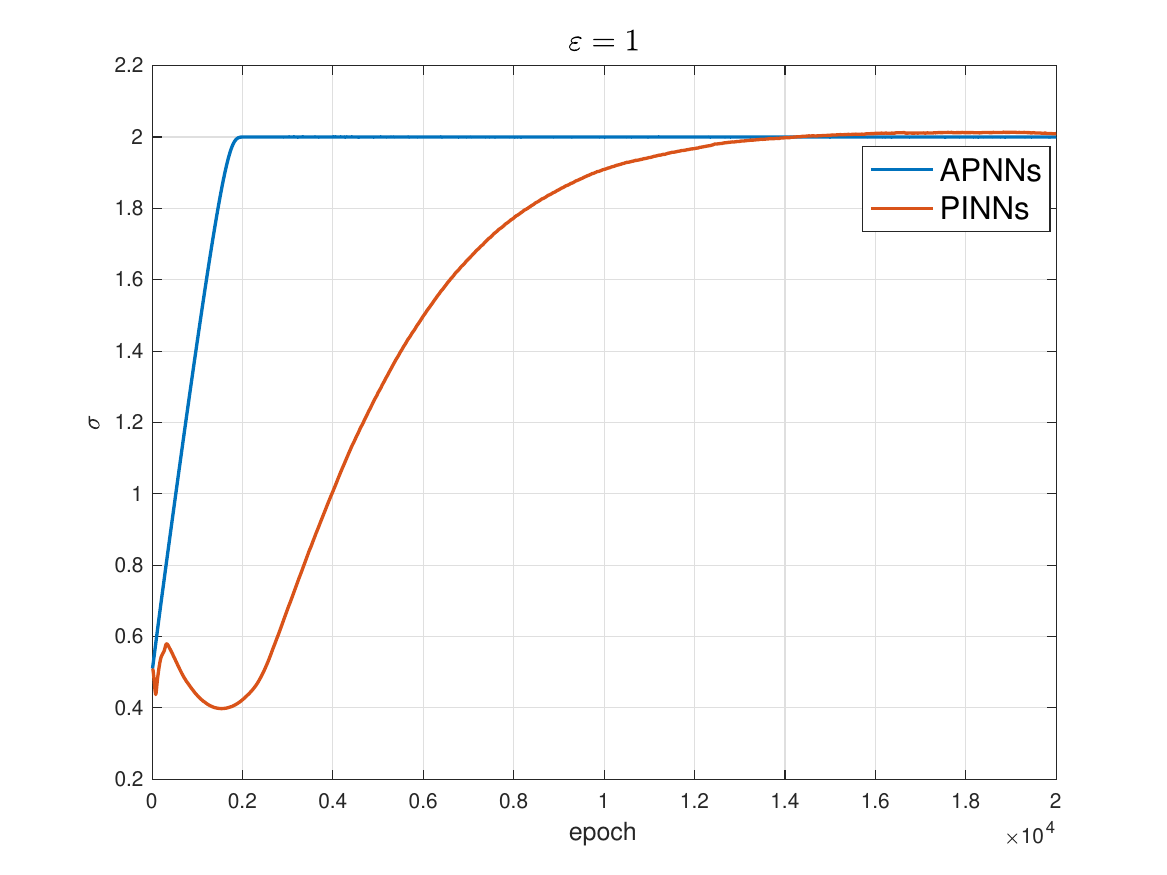}
    \includegraphics[width=0.49\textwidth]
    {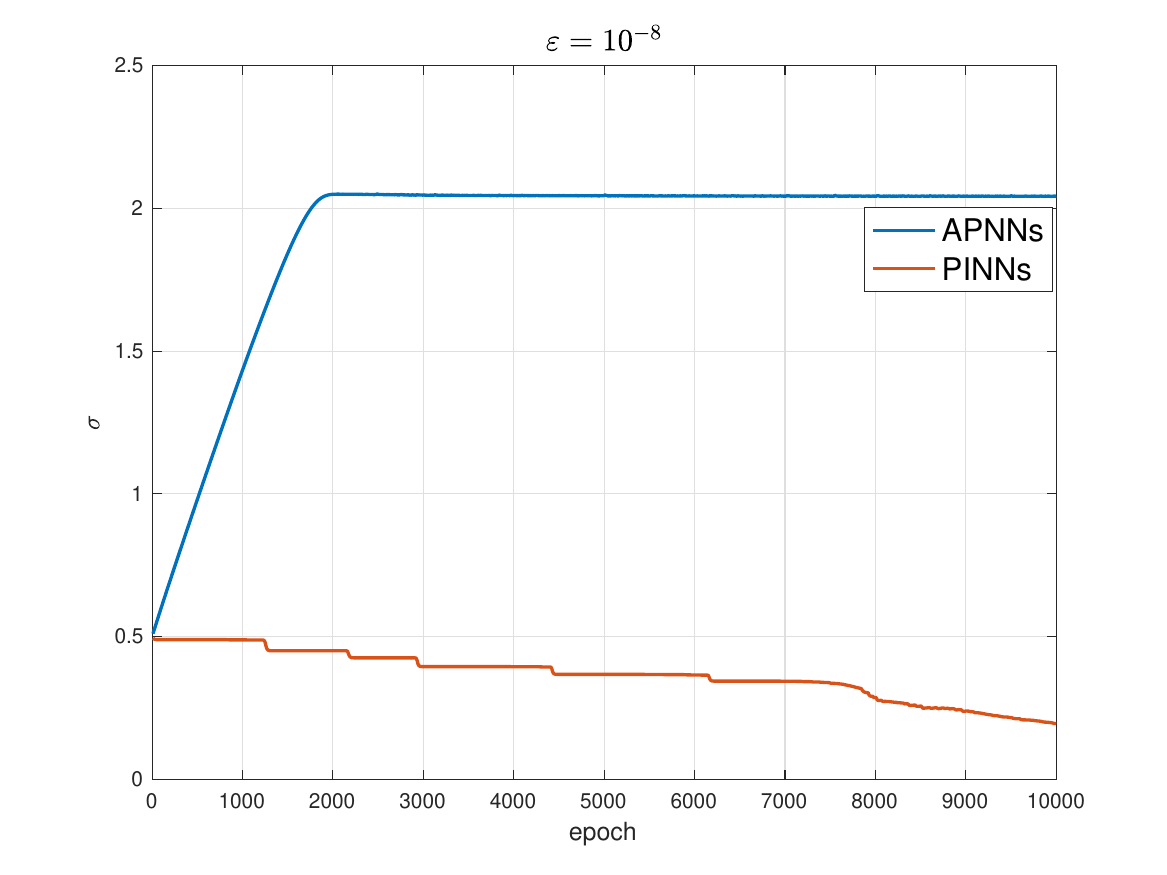}
\caption{Problem I with $\e=1$ and $\e=10^{-8}$: Inverse problem. Convergence of the target parameter $\sigma = 2$ with respect to epochs using APNNs and the standard PINNs with full observation data. }
    \label{fig:inverse_ep0.1}
\end{figure}

Figure \ref{fig:inverse_ep0.1} plots the convergence of the target parameter $\sigma=2$ for both APNNs and PINNs formulation. When $\e=1$, after training $20000$ epochs, both APNNs and PINNs can converge to the target $\sigma$ with the relative error be $1.20 \times 10^{-4}$ and $4.81 \times 10^{-3}$ respectively. When $\e=10^{-8}$, after training $10000$ epochs, a fast convergence is reached in APNNs, with the initial guess $\sigma_0 = 0.5$ and a final relative error be $O(10^{-2})$.
However, the regular PINNs completely fails
to recover the correct value of $\sigma$. 

\begin{figure}[H]
    \centering
    \includegraphics[width=0.6\textwidth]
    {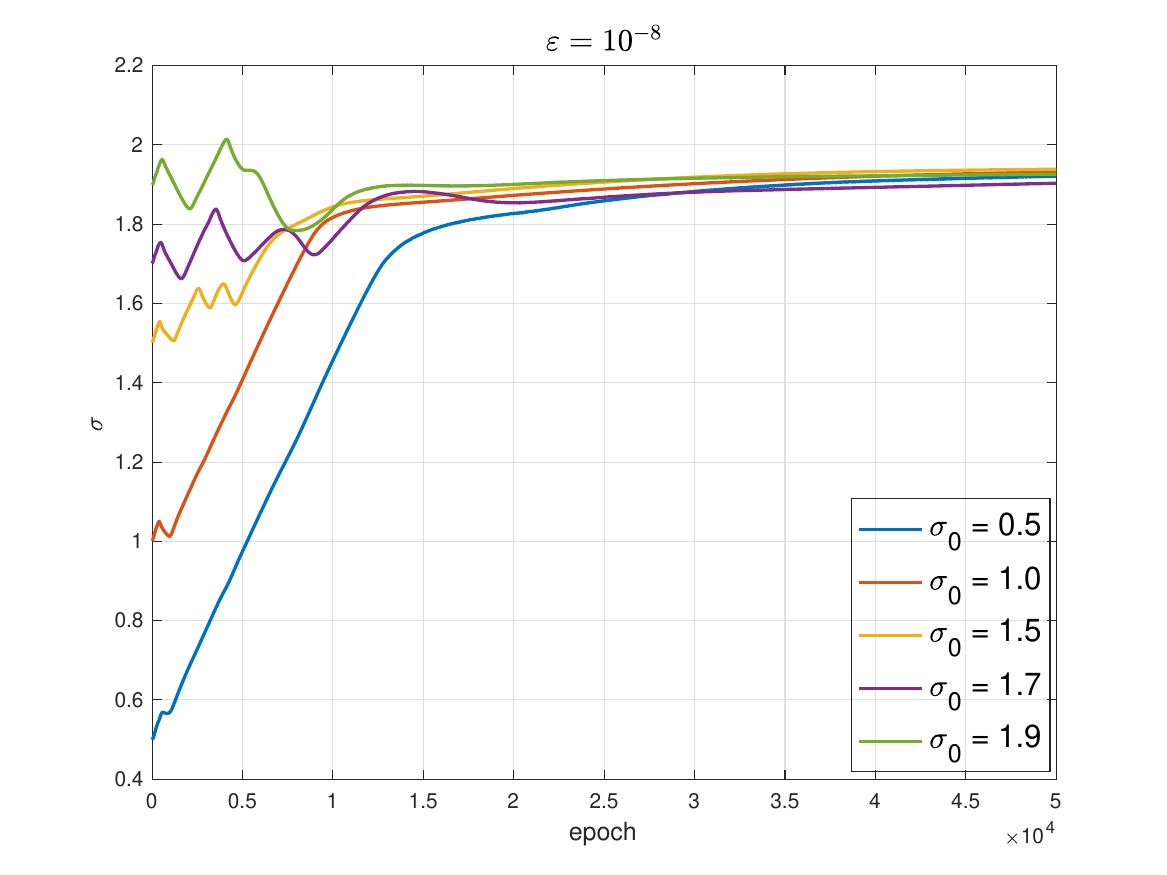}
\caption{Problem I with $\e=10^{-8}$: inverse problem with partial observation data. Convergence of the target parameter $\sigma$ with respect to epochs under different initial guesses $\sigma_0$ and learning rate $lr = 10^{-4}$ for \eqref{MM}. }
 \label{fig:inverse_partial}
\end{figure}

\noindent {\bf Scenario 2: Partially observated data. }  To further show the performance of APNNs, we examine a more practical and challenging setting where only partial observation data is available. Since $g$ is difficult to measure in real applications, we assume that only the measurement for $\rho$ can be accessed. Since PINNs method can not infer $\sigma$ when $\e$ is small even using full observation data, here we only test APNNs' performance on inferring target $\sigma$ when $\e$ is small. We use $100$ random measurements for  $\rho(t_d^i,x_d^i$) to estimate the targeted $\sigma=2$ under different initial guesses $\sigma_0 = 0.5, 1.0, 1.5, 1.7. 1.9$ under $\varepsilon = 10^{-8}$. The convergence of the target parameter $\sigma$ by APNNs can be found in Figure \ref{fig:inverse_partial}. Table \ref{tab:partial_observation_error} shows the relative errors between the target $\sigma$ and $\hat{\sigma}$ estimated by APNNs. The relative errors are around $3 \%$--$5 \%$ after training $50000$ epochs. For this example,  APNNs can obtain relatively acceptable results even when only partial data is provided, no matter how we choose the initial guess. The results are shown in Figure \ref{fig:inverse_partial}. The relative errors are around $1 \%$--$2 \%$ after training $50000$ epochs. The detailed results could be found in Table \ref{tab:partial_observation_error}.

\begin{table}[htbp]
\centering
\begin{tabular}{|c|c|c|c|c|c|}
\hline
$\mathbf{\sigma_0}$ & 0.5 & 1.0 & 1.5 & 1.7 & 1.9 \\
\hline
\textbf{Micro-Macro} &$3.94\times10^{-2}$  &$3.41 \times 10^{-2}$  & $\mathbf{3.07 \times 10^{-2}}$ & $4.84 \times 10^{-2}$ & $3.69 \times 10^{-2}$\\
\hline
\end{tabular}
\caption{Problem I with $\e = 10^{-8}$. Relative errors for the inverse problem with partially observed data for Micro-Macro Model \eqref{MM}. }
\label{tab:partial_observation_error}
\end{table}

%------------------------------------------------------
\subsection{Problem II: The Boltzmann-Poisson System}

In the second test, we consider the Boltzmann-Poisson system \eqref{eqn:BP}. Assume the incoming boundary conditions given by 
\begin{equation}\label{BC}
f(t,x,v)\Big|_{x_L} = F_L(v), \qquad
f(t,x,-v)\Big|_{x_R} = F_R(v), \quad \text{ for } v>0, 
\end{equation}
and Maxwellian function as the initial data
$ f(x,v,t=0)= M(v) $.  
Let the applied bias voltage $V=5$, $\beta=0.002$ and doping profile $c(x)$ be given by
$$ c(x) = 1 - (1-m)\rho(0,t=0)\left[ \tanh(\frac{x-0.3}{0.02}) - \tanh(\frac{x-0.7}{0.02})\right], $$
with $m = (1-0.001)/2$.  We set $\Delta x = 0.01$ and $\Delta t = 0.005$. 

In this example, besides $\rho(t_i,x_i), g(t_i,x_i,v_i)$ and $f(t_i,x_i,v_i)$, which are approximated by the same neural networks in the previous example, we approximate the extra latent solution $\phi(t_i,x_i)$ by one 16-layer deep neural network with $14$ hidden layers using the empirical loss of \eqref{empirical_Loss-PINN} and \eqref{empirical_Loss-APNN} for PINNs and APNNs, respectively. 

\begin{figure}[H]
    \centering
    \includegraphics[scale=0.425]
    {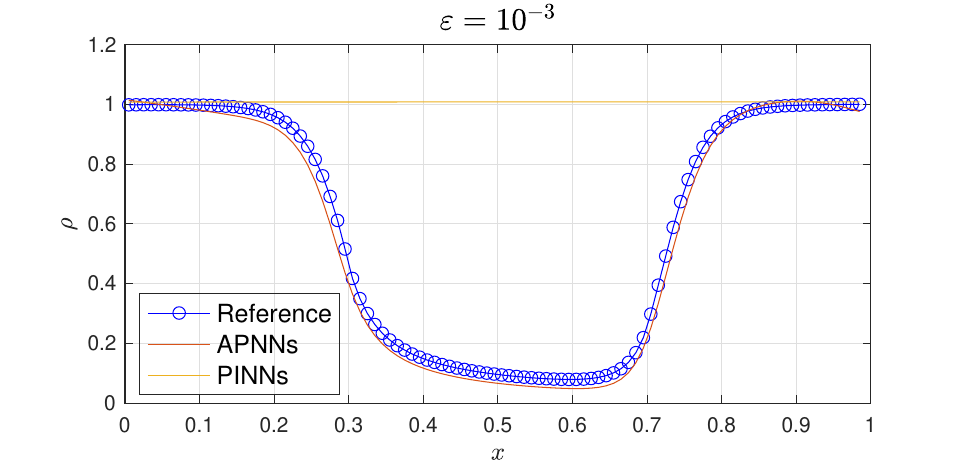}
     \includegraphics[scale=0.425]
    {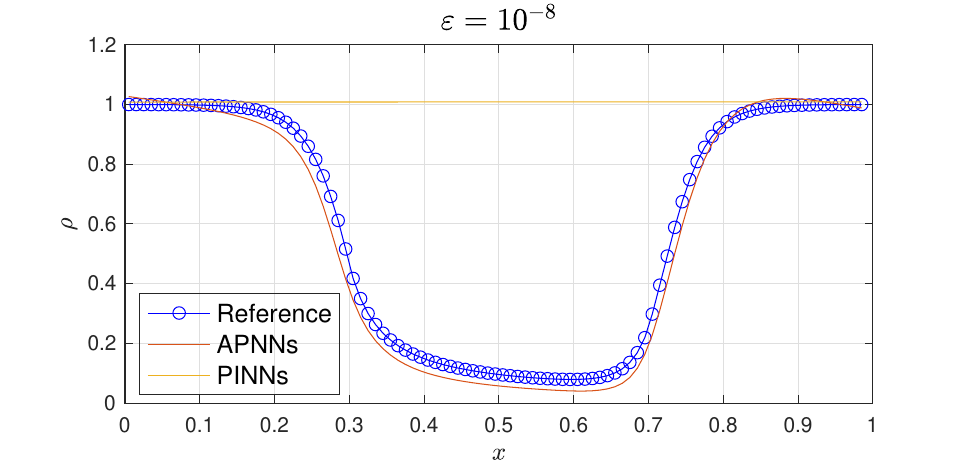}
\caption{Problems II with $\e = 10^{-3}$ and $\e = 10^{-8}$. Plot of density $\rho$ of PINNs, APNNs, and reference solutions at $T = 0.1$. }
\label{fig:BP_rho}
\end{figure}

\begin{figure}[H]
    \centering
    \includegraphics[scale=0.42]
    {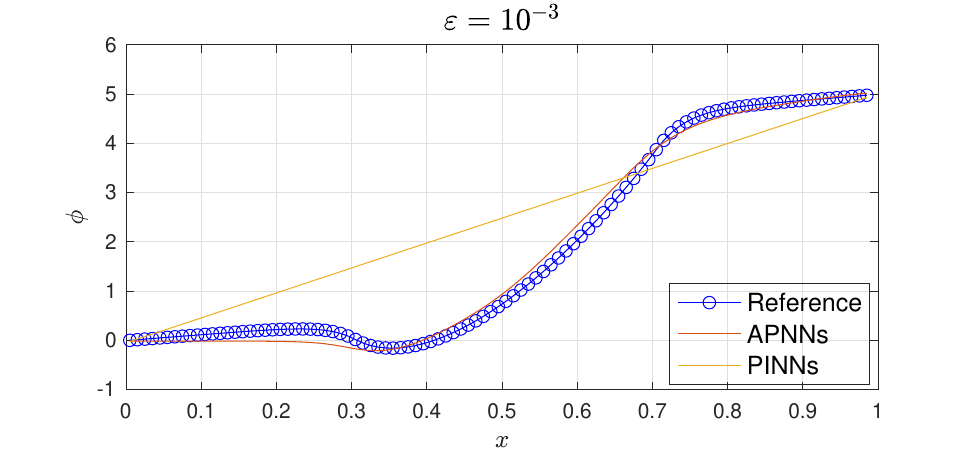}
     \includegraphics[scale=0.42]
    {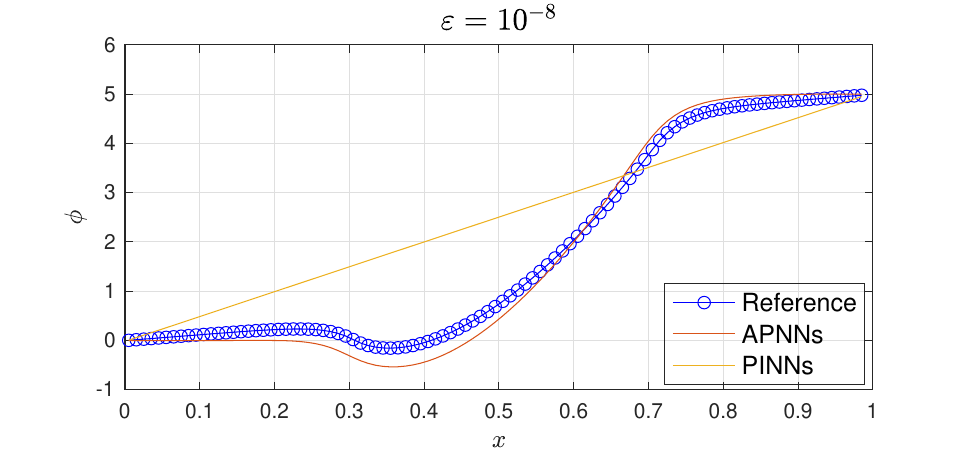}
\caption{Problems II with $\e = 10^{-3}$ and $\e = 10^{-8}$. Density $\phi$ of PINNs, APNNs and reference solutions at $T = 0.1$. }
\label{fig:BP_phi}
\end{figure}

The results for $\rho$ and $\phi$ are shown in Figure \ref{fig:BP_rho} and \ref{fig:BP_phi}, where we plot the solutions at final time $T=0.1$ by the AP deterministic solver \cite{JP2000}, PINNs and APNNs respectively. We find that APNNs solutions of $\rho$ and $\phi$ both match well with the reference solutions even when $\varepsilon$ is quite small. In contrast, the density by  PINNs converges to a trivial solution, which is a constant in our problem setup. In addition, the potential $\phi$ approximated by PINNs is completely off, which converges to a linear function such that only the Dirichlet boundary conditions are satisfied. 

Besides, we compute the resulting relative $\ell^2$ errors for $\rho$ and $\phi$ generated by PINNs and APNNs for $\e = 10^{-3}$ and $\e = 10^{-8}$ in Table \ref{tab:BP_comparison}. One observes that the relative $\ell^2$ errors of $\rho$ and $\phi$ for APNNs when $\e=10^{-3}$ are both of $O(10^{-2})$, whereas PINNs give a big error of $O(10^{-1})$ and completely wrong solution as shown in the above Figure \ref{fig:BP_rho} and Figure \ref{fig:BP_phi}. 

\begin{table}[h]
    \centering
    \begin{tabular}{|c|c|c|c|c|}
        \hline
        %\multirow{}{}{} 
        & \multicolumn{2}{c|}{$\e = 10^{-3}$} & \multicolumn{2}{c|}{$\e = 10^{-8}$} \\ \cline{2-5}
                          & PINNs & APNNs & PINNs & APNNs \\ \hline
        $\rho$            & $8.01 \times 10^{-1}$ & $ \mathbf{4.95 \times 10^{-2}}$ & $8.00 \times 10^{-1}$ & $\mathbf{6.56 \times 10^{-2}}$ \\ \hline
        $\phi$            & $4.05 \times 10^{-1}$ & $ \mathbf{6.32 \times 10^{-2}}$ & $4.04 \times 10^{-1}$ & $\mathbf{7.66 \times 10^{-2}}$ \\ \hline
    \end{tabular}
    \caption{Comparison of $\rho$ and $\phi$ under two different $\epsilon$ values using two different methods}
    \label{tab:BP_comparison}
\end{table}

\bigskip

\noindent \textbf{Inverse Problem.} 

We further consider another inverse problem inferring the scattering coefficient from available measurement data using the APNNs formulation in BP system. Similarly, we consider using both APNNs and PINNs in the first scenario, where full observation data is available. We consider only APNNs in the second scenario since PINNs failed even with full observation data.

\bigskip

\noindent {\bf Scenario 1: Full observation data. } 
First, we consider an ideal inverse problem setup inferring the scattering coefficient $\sigma$ by assuming the availability of a limited number of the full observation dataset. We obtain synthetic dataset for $\rho(t_d^i,x_d^i)$, $f(t_d^i,x_d^i,v_d^i)$, $g(t_d^i,x_d^i,v_d^i)$ and $\phi(t_d^i,x_d^i)$, where  $g(t_d^i,x_d^i,v_d^i)$ is obtained in the way similar to Problem I. Besides the APNNs loss \eqref{Inverse_APNN} and PINNs loss \eqref{Inverse_PINN}, we need an extra term $L_{{\phi}_{{d}}}^{APNN}(\theta)$, here
\begin{equation}
    L_{{\phi}_{{d}}}^{APNN}(\theta) = \frac{\omega_d^\phi}{N_{d_4}}\sum_{i=1}^{N_{d_4}}\left|\phi^{NN}(t_d^i,x_d^i;\theta)-\phi(t_d^i,x_d^i)\right|^2. 
    \label{Inverse_phi}
\end{equation}
where $\omega_d^\phi = 1$, and $N_{d_4}$ is the number of data points for $\phi$.
We train the network model on measurements with $N_{d_1}, N_{d_4} = 100$ samples randomly selected in the domain $(t,x) \in [0,0.1] \times [0,1]$, and $N_{d_2}, N_{d_3} = 100$ are samples randomly selected in the domain $(t,x,v) \in [0,0.1] \times [0,1] \times \Omega_v$, where $\Omega_v$ is the discrete points in the velocity space obtained by the Hermite quadrature rule from which $20 \%$ points are randomly selected for validation.

\begin{figure}[H]
    \centering
    \includegraphics[width=0.49\textwidth]
{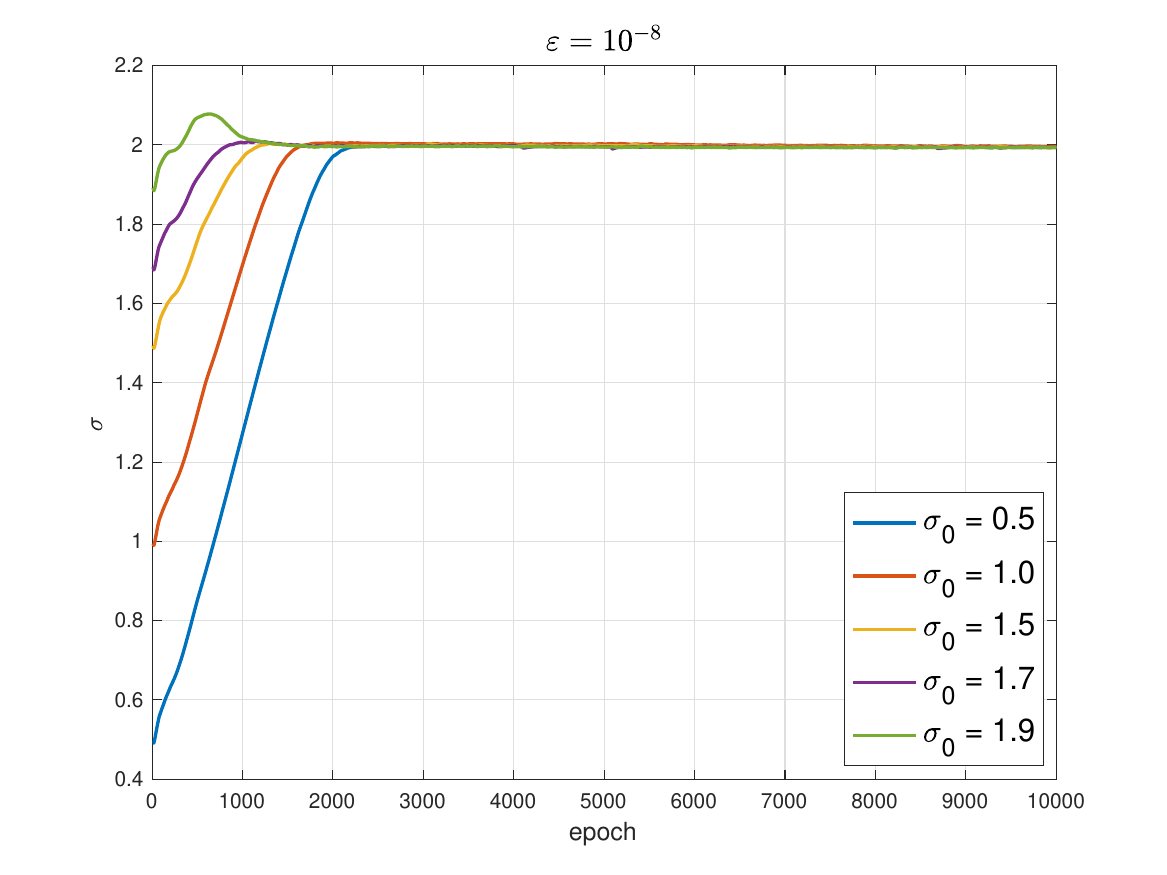}
    \includegraphics[width=0.49\textwidth]
    {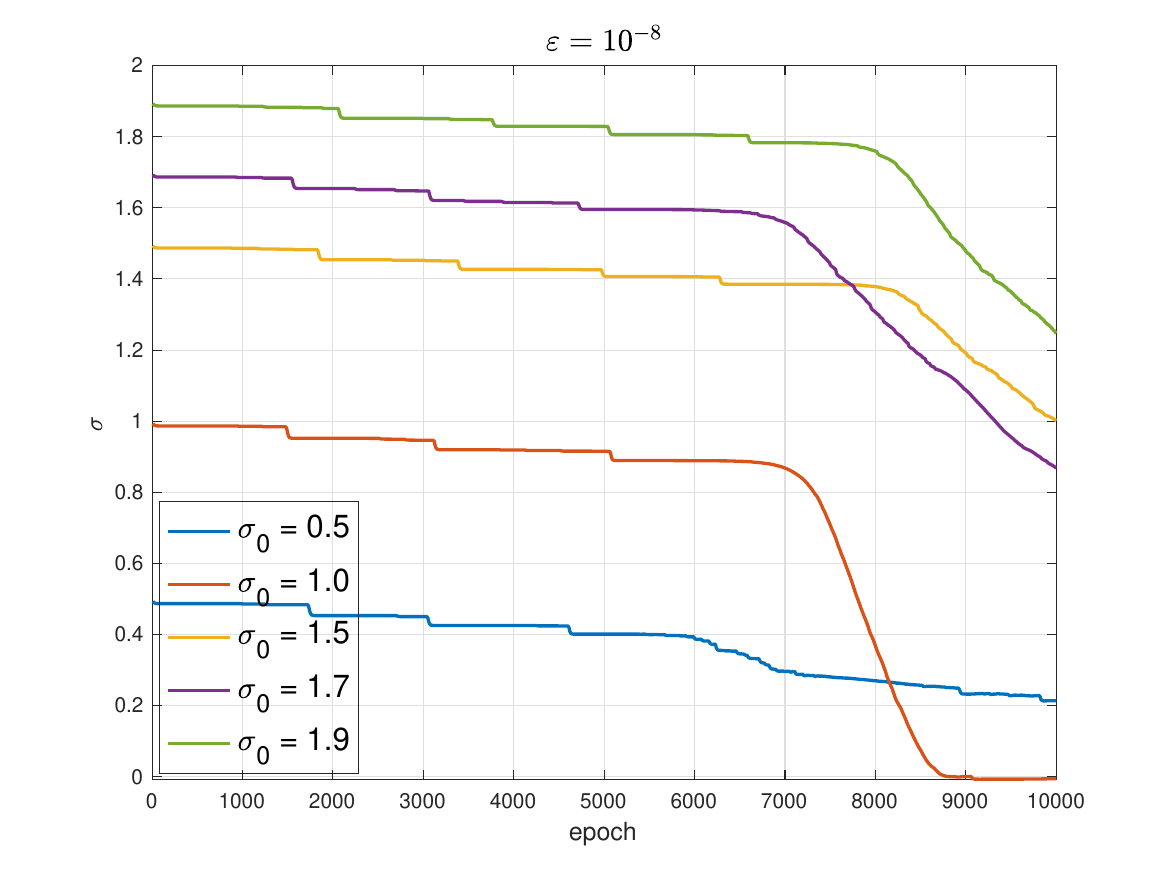}
\caption{Problem II with $\e=10^{-8}$: Inverse problem with full observation data. Convergence of the target parameter $\sigma$ with respect to epochs using APNNs (left)and PINNs (right) under different initial guesses $\sigma_0$.}
    \label{fig:test2_inverse_full}
\end{figure}

Figure \ref{fig:test2_inverse_full} plots the convergence of the target parameter $\sigma=2$ for both APNNs and PINNs formulation. On the left of the figure, we can find when $\e=10^{-8}$, after training $10000$ epochs, a fast convergence is reached in APNNs, with different initial guesses $\sigma_0 = 0.5, 1.0, 1.5, 1.7. 1.9$. Final relative errors are $O(10^{-3})$ shown in Table \ref{tab:test2_full_observation_error}. On the right of the figure, we can find the estimated $\hat{\sigma}$ by PINNs can not converge to the target $\sigma = 2$ under different initial guesses $\sigma_0$. The final relative errors are around $O(10^{-1})$ for PINNs. This can further emphasize the advantages of our APNNs method.

\begin{table}[htbp]
\centering
\begin{tabular}{|c|c|c|c|c|c|}
\hline
$\mathbf{\sigma_0}$ & 0.5 & 1.0 & 1.5 & 1.7 & 1.9 \\
\hline
\textbf{PINNs} &$8.93 \times 10^{-1}$  &$1.01 \times 10^{0}$ & $5.64 \times 10^{-1}$ & $4.99 \times 10^{-1}$ & $3.75 \times  10^{-1}$\\
\hline
\textbf{APNNs} &$\mathbf{3.07 \times 10^{-3}}$  &$ \mathbf{1.85 \times 10^{-3}}$ & $ \mathbf{2.95 \times 10^{-3}}$ & $ \mathbf{3.18 \times 10^{-3}}$ & $ \mathbf{3.22 \times 10^{-3}}$\\
\hline
\end{tabular}
\caption{Problem II with $\e = 10^{-8}$. Relative errors of PINNs and APNNs for the inverse problem with full observation under different initial guess $\sigma_0$. }
\label{tab:test2_full_observation_error}
\end{table}

\noindent {\bf Scenario 2: Partial observation data. }
Since PINNs with full observation data have been shown to fail to infer $\sigma$ when $\e$ is small. In this problem, we only consider using APNNs to infer $\sigma = 2$ in \eqref{eqn:BP} under an extremely small $\e = 10^{-8}$ to further show the performance of APNNs in BP system. We examine a more practical and challenging setting where only partial observation data is available. Since the potential $\phi$ and microscopic quantity $g$ are difficult to measure in real applications, we assume that only the measurement for $\rho$ can be accessed. Similar to Problem I, we use $100$ random measurements for  $\rho(t_d^i,x_d^i)$ and  $\phi(t_d^i,x_d^i)$ to estimate the targeted $\sigma=2$ under different initial guesses $\sigma_0 = 0.5, 1.0, 1.5, 1.7. 1.9$ under $\varepsilon = 10^{-8}$. 
For the problem \eqref{MM}, Table \ref{tab:BP_partial} shows the estimated $\sigma$ after training $50000$ epochs. The relative errors are around $1 \%$--$2 \%$ after training $50000$ epochs. 

\begin{table}[htbp]
\centering
\begin{tabular}{|c|c|c|c|c|c|}
\hline
$\mathbf{\sigma_0}$ & 0.5 & 1.0 & 1.5 & 1.7 & 1.9 \\
\hline
\small{\textbf{Micro-Macro BP}} &$2.70 \times 10^{-2}$  &$2.58 \times 10^{-2}$ & $2.27 \times 10^{-2}$ & $1.25 \times 10^{-2}$ & $\mathbf{5.40 \times 10^{-3}}$\\
\hline
\end{tabular}
\caption{Problem II with $\e = 10^{-8}$. Relative errors for the inverse problem with partial observation for Micro-Macro Model \eqref{MM}. 
}
\label{tab:BP_partial}
\end{table}

%------------------------------------------------------

\section{Conclusions and future work}

In this paper, we construct an AP neural network in the micro-macro decomposition framework for the semiconductor Boltzmann and the Boltzmann-Poisson system with multiple scales. Based on the Universal Approximation Theorem and the hypocoercivity tools, convergence analysis for both the loss function and NN approximated solution are conducted. Several numerical tests are carried out to demonstrate the performance of the APNNs, both for inverse and forward problems with multiple scaling. 

In the future work, we will further improve our analysis result by studying the Barron-type functions \cite{Gu-Ng} and posterior estimates of the solution error. In addition, we shall study higher dimensional problems and extend to the uncertainty quantification problems with more practical applications. 

\section*{Acknowledgement}
\label{sec:ack}
L.~Liu acknowledges the support by National Key R\&D Program of China (2021YFA1001200), Ministry of Science and Technology in China, Early Career Scheme (24301021) and General Research Fund (14303022 \& 14301423) funded by Research Grants Council of Hong Kong from 2021-2023.  Y.~Wang acknowledges the support by National Natural Science Foundation of China (grant No. 12301559).

\section*{Appendix}

\section*{Velocity discretizations. }
For completeness, we briefly mention the velocity discretization similar to what has been studied in \cite{JP00}. Set $f(t,x,v) = \psi(t,x,v) M(v)$, where $M(v) = \frac{1}{\sqrt{\pi}} e^{-v^2}$, with 
\begin{equation}\label{Psi} \psi(t,x,v) = \sum_{k=0}^N \psi_k(t,x) \tilde H_k(v), \end{equation}
being the Hermite expansion. 
For notation simplicity, we omit the $t$ and $x$ dependence of functions below. 
Here $\tilde H_k$ are the renormalized Hermite polynomials
defined as $\tilde H_{-1}=0$, $\tilde H_0 = 1/\pi^{1/4}$ and 
$$ \tilde H_{j+1} = v \sqrt{\frac{2}{j+1}}\tilde H_j - \sqrt{\frac{j}{j+1}}\tilde H_{j-1} \quad \text{for } j\geq 0, $$
satisfying $\partial_v \tilde H_j = \sqrt{2 j}\, \tilde H_{j-1}$. 
The inverse Hermite expansion is given by 
\begin{equation}\label{I-Psi} \psi_k = \sum_{j=0}^N \psi(v_j)\, \tilde H_k(v_j)\, w_j, 
\end{equation}
where $(v_j, w_j)$ are the points and corresponding weights of the Gauss-Hermite quadrature rule. 
Thus the collision operator $Q$ in \eqref{Boltz-eqn} can be computed by
$$ Q(f)(v) = M(v) \sum_{j=0}^N \sigma(v, v_j)\, \psi(v_j)\, w_j - \lambda(v) f(v), $$
with $\lambda(v) = \sum_{j=0}^N \sigma(v, v_j)\, w_j$. From \eqref{Psi} and \eqref{I-Psi}, one computes the derivative in $v$ by
\begin{equation*}
\begin{aligned}
\partial_v \psi & = \sum_{k=0}^N \psi_k\, \partial_v \tilde H_k(v) 
= \sum_{k=0}^N \psi_k \sqrt{2k}\, \tilde H_{k-1}(v) \\
& = \sum_{k=0}^N \sum_{j=0}^N \psi(v_j) \, \tilde H_k(v_j) w_j \sqrt{2k}\, \tilde H_{k-1}(v) 
 = \sum_{j=0}^N \psi(v_j)\, C_j(v),
\end{aligned}
\end{equation*}
where $C_j(v) = \sum_{k=0}^N \sqrt{2k}\, \tilde H_k(v_j) \tilde H_{k-1}(v) w_j$, 
and can be precomputed before the time iteration.

\section*{Loss functions for PINNs and APNNs}
The empirical risk for PINN is as follows:
\begin{equation}
\label{empirical_Loss-PINN}
\small{\begin{aligned}
 \mathcal{R}_{\mathrm{PINN}}^{\e} & = 
 \frac{1}{N_1} \sum_{i=1}^{N_1} \Big| \e \partial_t f_\theta^{\mathrm{NN}}(t_i,x_i,v_i) +  \bv \partial_x f_\theta^{\mathrm{NN}}(t_i,x_i,v_i)
+ \partial_x \phi(t_i,x_i)\, \partial_v f_\theta^{\mathrm{NN}}(t_i,x_i,v_i) - \frac{1}{\e}\mathcal{Q}(f_\theta^{\mathrm{NN}}(t_i,x_i,v_i))|^2 \,\\[6pt]
& + \frac{\lambda_1}{N_2} \sum_{i=1}^{N_2} \left|\mathcal{B} (f_{\theta}^{\mathrm{NN}}(t_i,x_i,v_i)) - f_{\text{BC}}(t_i,x_i,v_i)\right|^2  + \frac{\lambda_2}{N_3} \sum_{i=1}^{N_3} \left | \mathcal{I}(f_\theta^{\mathrm{NN}}(t_i,x_i,v_i))- f_{\text{IC}}(t_i,x_i,v_i) \right |^2. 
\end{aligned}}
\end{equation}
where $N_1$, $N_2$, $N_3$ are the number of sample points of $\mathcal{T} \times \mathcal{D} \times \Omega$, $\mathcal{T} \times \partial \mathcal{D} \times \Omega$ and $\mathcal{D} \times \Omega$. 
The empirical risk for APNN is as follows:
\eqref{Macro}--\eqref{Micro} as the APNN loss function: 
\begin{equation}
\label{empirical_Loss-APNN}
\small{\begin{aligned}
\mathcal{R}_{\mathrm{APNN}}^{\e} = & \frac{1}{N_1} \sum_{i=1}^{N_1}\left|\partial_t \rho_\theta^{\mathrm{NN}}(t_i,x_i)+\nabla_{\bx} \cdot\left\langle\bv g_\theta^{\mathrm{NN}}(t_i,x_i)\right\rangle + \red{ \nabla_{\bx}\phi(t_i,x_i) \cdot  \left\langle \nabla_{\bv} g_\theta^{\mathrm{NN}}(t_i,x_i)\right\rangle } \right|^2 \,\\[6pt]
& + \frac{1}{N_2} \sum_{i=1}^{N_2} \Big| \e^2 \partial_t g_\theta^{\mathrm{NN}}(t_i,x_i,v_i)
 + \e(I-\Pi) \left(\bv \cdot \nabla_{\bx} g_\theta^{\mathrm{NN}}(t_i,x_i,v_i) + \red{\nabla_{\bx}\phi(t_i,x_i) \cdot \nabla_{\bv} g_\theta^{\mathrm{NN}}(t_i,x_i,v_i) }\right) \\[6pt]
&   - 2\bv \cdot \nabla_{\bx}\phi(t_i,x_i)\, \rho_\theta^{\mathrm{NN}}(t_i,x_i)\, M(v)  
 + \bv \cdot \nabla_{\bx} \rho_\theta^{\mathrm{NN}}(t_i,x_i) M(v)  - \mathcal{Q}( g_\theta^{\mathrm{NN}}(t_i,x_i,v_i) ) \Big|^2 \,\\[6pt]
& + \frac{\lambda_1}{N_3} \left|\mathcal{B}  \left(\rho_\theta^{\mathrm{NN}}(t_i,x_i) M(v) + \e g_\theta^{\mathrm{NN}}(t_i,x_i,v_i)\right) - f_{\text{BC}}(t_i,x_i,v_i)\right|^2 \,\\[6pt]
 &+\frac{\lambda_2}{N_4} \left|\mathcal{I}\left(\rho_\theta^{\mathrm{NN}}(t_i,x_i) M(v) +\e g_\theta^{\mathrm{NN}}(t_i,x_i,v_i)\right)- f_{\text{IC}}(t_i,x_i,v_i) \right|^2, 
\end{aligned}}
\end{equation}
where $N_1$,$N_2$, $N_3$, $N_4$ are the number of sample points of $\mathcal{T} \times \mathcal{D}$, $\mathcal{T} \times \mathcal{D} \times \Omega$, $\mathcal{T} \times \partial \mathcal{D} \times \Omega$ and $\mathcal{D} \times \Omega$.

Regarding the incoming boundary condition given as
$$ f(t,x,v)\Big|_{x_L} = F_L(v), \qquad
f(t,x,-v)\Big|_{x_R} = F_R(v), \quad \text{ for } v>0, 
$$
we look at the third term of \eqref{Loss-APNN}, with the discretized form of the integral shown by
\begin{equation*}
\begin{aligned}
&\sum_{i}\sum_{j \text{ for }v_j>0} 
\left| \rho_\theta^{\mathrm{NN}}(t_i, x_L) M(v_j) + 
\e g_\theta^{\mathrm{NN}}(t_i, x_L, v_j) - F_L(v_j) \right|^2 w_j \, \Delta t \\
& + \sum_{i}\sum_{j \text{ for }v_j<0}
\left| \rho_\theta^{\mathrm{NN}}(t_i, x_R) M(v_j) + 
\e g_\theta^{\mathrm{NN}}(t_i, x_R, v_j) - F_R(v_j) \right|^2 w_j\, \Delta t. 
\end{aligned}
\end{equation*}

For the integrals in losses of PINNs and APNNs, the operator $\langle \cdot \rangle$ in  \eqref{bracket} and $\Pi(\cdot)$ in \eqref{Ans} are computed by the quadrature rules.

%------------------------------------------------------

\bibliographystyle{siam}
\bibliography{Submission_main_clean.bib}
\end{document}